\documentclass[iop]{emulateapj}

\usepackage{amssymb,amsmath,amsfonts,amsbsy}
\usepackage{color}
\usepackage{natbib}
\usepackage{enumerate}
\usepackage{graphicx}

\begin{document}

\title{Minkowski Tensors in Two Dimensions - Probing the Morphology and Isotropy of the Matter and Galaxy Density Fields}

\author{Stephen Appleby$^{a}$}\email{stephen@kias.re.kr}
\author{Pravabati Chingangbam$^{b}$}
\author{Changbom Park$^{a}$}
\author{Sungwook E. Hong$^{c}$}
\author{Juhan Kim$^{d}$}
\author{Vidhya Ganesan$^{b,e}$}
\affiliation{$^{a}$School of Physics, Korea Institute for Advanced Study, 85
Hoegiro, Dongdaemun-gu, Seoul 02455, Korea}
\affiliation{$^{b}$Indian Institute of Astrophysics, Koramangala II Block, Bangalore 560 034, India}
\affiliation{$^{c}$Korea Astronomy and Space Science Institute, 776 Daedeokdae-ro, Yuseong-gu, Daejeon 34055, Korea}
\affiliation{$^{d}$Center for Advanced Computation, Korea Institute for Advanced Study, 85 Hoegiro, Dongdaemun-gu, Seoul 02455, Korea}
\affiliation{$^{e}$Indian Institute of Science, Bangalore 560 034, India}

\begin{abstract}
We apply the Minkowski Tensor statistics to two dimensional slices of the three dimensional matter density field. The Minkowski Tensors are a set of functions that are sensitive to directionally dependent signals in the data, and furthermore can be used to quantify the mean shape of density fields. We begin by reviewing the definition of Minkowski Tensors and introducing a method of calculating them from a discretely sampled field. Focusing on the statistic $W^{1,1}_{2}$ -- a $2 \times 2$ matrix -- we calculate its value for both the entire excursion set and for individual connected regions and holes within the set. To study the morphology of structures within the excursion set, we calculate the eigenvalues $\lambda_{1},\lambda_{2}$ for the matrix $W^{1,1}_{2}$ of each distinct connected region and hole and measure their mean shape using the ratio $\beta \equiv \langle \lambda_{2}/\lambda_{1} \rangle$. We compare both $W^{1,1}_{2}$ and $\beta$ for a Gaussian field and a smoothed density field generated from the latest Horizon Run 4 cosmological simulation, to study the effect of gravitational collapse on these functions. The global statistic $W^{1,1}_{2}$ is essentially independent of gravitational collapse, as the process maintains statistical isotropy. However, $\beta$ is modified significantly, with overdensities becoming relatively more circular compared to underdensities at low redshifts. When applying the statistics to a redshift-space distorted density field, the matrix $W^{1,1}_{2}$ is no longer proportional to the identity matrix and measurements of its diagonal elements can be used to probe the large-scale velocity field. 
\end{abstract}

\maketitle

\section{\label{sec:1}Introduction}

One of the fundamental axioms implicit within the standard cosmological model is that the distribution of matter in the Universe is statistically isotropic and homogeneous when smoothed over suitably large scales. This condition is very well observed in the early epoch of radiation and matter domination, where fluctuations in the dark matter density field are small. However, the scale at which this remains true at low redshifts is less clear as non-linear gravitational evolution generates a complex web of structures. We expect alignment of structures due to their position within the cosmic web \citep{Lee:2007nq,1538-4357-567-2-L111,Aubert:2004mu,Codis:2014awa,1538-4357-652-2-L75, codis_dubois_pichon_devriendt_slyz_2014, doi:10.1093/mnras/stv1570, Paz:2008na, Hahn:2007ui}, and a bias in the clustering properties of galaxies. The dark matter field exhibits coherent structures even at very large scales ($\sim 100 h^{-1} \, {\rm Mpc}$), so the scale at which alignments cease to become significant remains an open question. 

Furthermore when introducing an observer, one can state that the observed distribution of dark matter tracers (typically galaxies) are neither homogeneous nor isotropic - selection effects generate a non-trivial radial profile in the observed number density, and redshift-space distortion effects will modify the apparent positions of galaxies along the line of sight. Line of sight effects will generate a bias in the detection of structures perpendicular to the line of sight. Isotropy of the galaxy sample is lost via masks and boundaries. If we can measure the degree of anisotropy in data sets, then we can potentially minimize observational systematics and in the case of redshift-space distortion, constrain the growth rate.

The $N+1$ Minkowski Functionals are a set of scalar quantities that characterize the morphology and topology of an $N$ dimensional field \citep{1970Ap......6..320D,Adler,Gott:1986uz,Hamilton:1986, Gott:1986uz,Ryden:1988rk,1989ApJ...340..625G,1989ApJ...345..618M,1992ApJ...387....1P,1991ApJ...378..457P, Matsubara:1994we,1996ApJ...457...13M,Schmalzing:1995qn,2005ApJ...633....1P,Kerscher:2001gm}. Since they are scalar quantities, they cannot inform us of any directionally dependent information contained within the data. The concept of Minkowski Functionals can be generalised to vector and tensor counterparts \citep{McMullen:1997,Alesker1999,2002LNP...600..238B,HugSchSch07,1367-2630-15-8-083028,JMI:JMI3331} - these quantities are typically defined as integrals of some higher rank quantity over the boundary of an excursion set. As such, they contain information not present in the standard Minkowski Functionals. In particular they can be used to identify globally anisotropic signals in the data, as well as provide a measure of the shape of peaks/troughs of a density field when applied to individual connected regions and holes in an excursion set. In both instances, the Minkowski Tensors measure directions associated with a boundary. 

The application of Minkowski Tensors to cosmology is a relatively new phenomenon. \citet{2017JCAP...06..023G} have applied a
Minkowski Tensor that encodes shape and alignment information of structures to the two dimensional CMB fields. The authors showed that the 2015 $E$-mode PLANCK data \citep{Adam:2015tpy} exhibits higher than $3\sigma$ level of anisotropy or alignment of hotspots and coldspots. Analytic expressions of translation invariant Minkowski Tensors for Gaussian random fields in two dimensions have been derived in \citet{Chingangbam:2017uqv}. 

In this work, we apply the Minkowski Tensors to two dimensional slices of the dark matter density field. We first review the generalisation of the Minkowski Functionals to their tensor equivalents. We then ask how these quantities can be used to test the isotropy of the field. Throughout this paper we focus on two dimensional slices of a three dimensional volume - in a companion paper we consider the three dimensional generalisation of these statistics. 

In the following section we provide a thorough explanation of our construction of the Minkowski Functionals and Tensors, by generating the boundary of an excursion set in two dimensions. We then define the Minkowski Tensors, and show how they can be calculated for a discrete field and bounding perimeter. We apply our algorithms to a Gaussian random field, connecting our numerical results to analytic predictions wherever possible. We close by applying the statistics to the late time gravitationally evolved dark matter field, using the latest Horizon Run simulation.

\section{\label{sec:2}Generating the boundary of an excursion set - two dimensions} 

We begin with a discussion of our construction of a bounding perimeter enclosing a subset of a density field in two dimensions. Our analysis in this section will closely follow that of \citet{JMI:JMI3331}, but we detail the method for completeness.

Our starting point is a discrete two dimensional density field $\delta_{ij}$ on a regular lattice spanned by $i,j$ subscripts, $1 \le i \le N_{\rm pix}$, $1 \le j \le N_{\rm pix}$ where $N_{\rm pix}$ is the number of grid points in one dimension. The domain is chosen to be a square with periodic boundary conditions but this condition is not necessary. We define a dimensionless density threshold $\nu = \delta_{\rm c}/\sigma_{0}$, where $\delta_{\rm c}$ is a constant density threshold and $\sigma_{0}$ is the rms fluctuation of $\delta_{ij}$. A perimeter of constant density $\delta_{\rm c} = \sigma_{0}\nu$ defines an excursion set of the field. We can label each $(i,j)$ pixel as either `in' or `out' of the excursion set according to $\delta_{ij} > \nu \sigma_{0}$ or $\delta_{ij} < \nu \sigma_{0}$ respectively. Our intention is to construct a closed boundary perimeter that separates in/out pixels. 

\begin{figure*}
  \centering
  \includegraphics[width=0.95\textwidth]{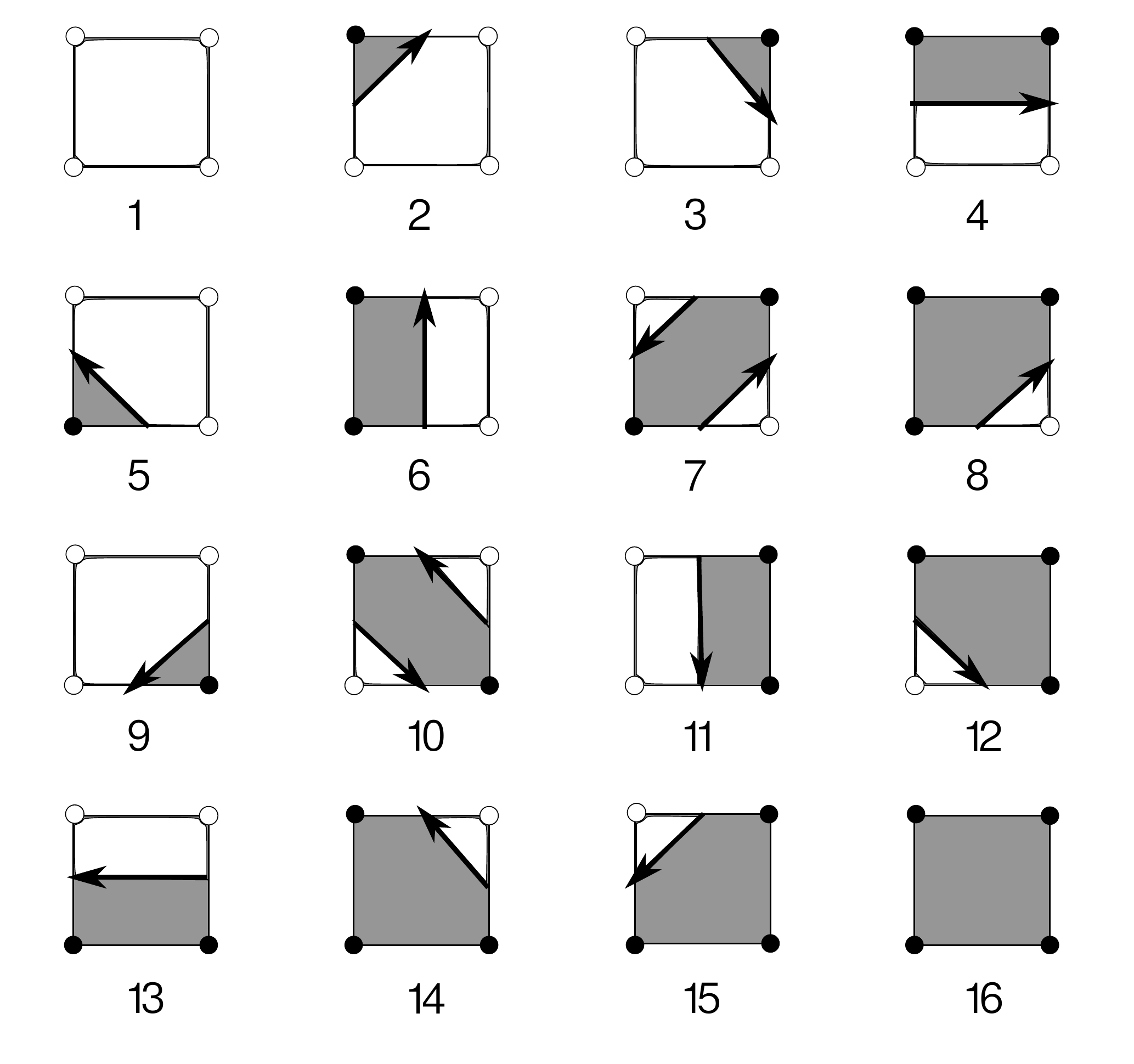}
  \caption{The marching squares algorithm - each set of four adjacent pixels for the discretized field $\delta_{ij}$ can take one of sixteen possible combinations of `in' or `out' states, here black circles denote points in our grid for which the density lies inside the excursion set $\delta_{ij} > \sigma_{0}\nu$ and white circles are `out' points with $\delta_{ij} < \sigma_{0}\nu$. We label each case with an integer $1 \le N_{\rm c} \le 16$. Between each `in' and `out' state we linearly interpolate between corners of the box to find the point along the edge of the square that satisfies $\delta = \sigma_{0}\nu$. We then connect these vertices, shown as solid black arrows - the arrow defines the boundary of the excursion region and is directed such that it always flows anti-clockwise around the `in' states. The cases $N_{\rm c} = 7$ and $N_{\rm c} = 10$ are ambiguous, as discussed in the text.   }
  \label{fig:1}
\end{figure*}

We adopt the method of marching squares \citep{1742-5468-2008-12-P12015}. The method performs a single sweep through the entire grid, systematically from one corner to the opposite. At each grid point $(i,j)$, we generate a square from its adjacent pixels - they are $(i,j), (i+1,j), (i,j+1), (i+1,j+1)$. Each of these pixels can either be `in' or `out' of the excursion set, so there are $2^{4}=16$ possible unique states of the square. In Figure \ref{fig:1} we exhibit the standard sixteen states, where we use the integer $1 \le N_{\rm c} \le 16$ to define each case as labeled. Each point denotes a $\delta_{ij}$ vertex, black are `in' states $\delta_{ij} > \sigma_{0}\nu$ and white `out' $\delta_{ij} < \sigma_{0}\nu$. A closed bounding perimeter is then constructed based on the sixteen cases, by linearly interpolating along the edges of the squares. Specifically, one can note that whenever an `in' and `out' state are joined along an edge of a square, we linearly interpolate between the values of $\delta$ at these vertices, along the edge, to the point at which $\delta = \sigma_{0}\nu$ is reached. This defines a vertex in the bounding perimeter. Vertices are then joined according to Figure \ref{fig:1} - this defines the edge components of the boundary which correspond to the solid arrows in the figure. Finally, we use trigonometry to calculate the area enclosed by the bounding perimeter in each square (the shaded region in each case in Figure \ref{fig:1}). The perimeter of the boundary, exhibited as black arrows in the figure, is directed such that the arrow always flows anti-clockwise around the `in' vertices $\delta_{ij} > \sigma_{0} \nu$.

\begin{figure}
  \includegraphics[width=0.45\textwidth]{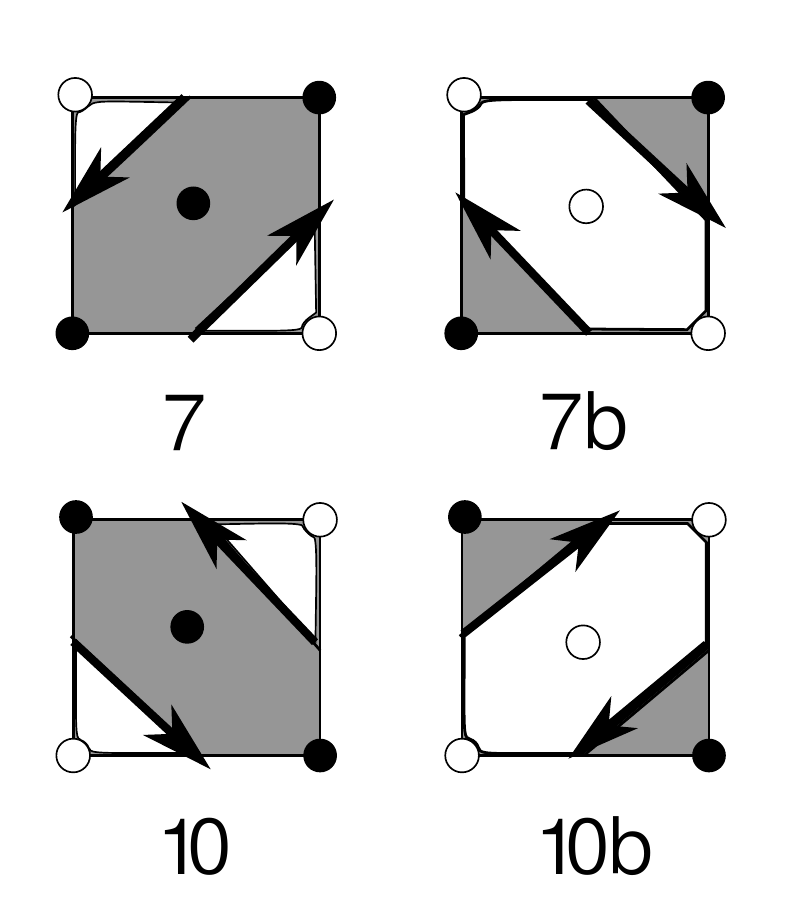}
  \caption{The ambiguous cases $7$ and $10$ - the `in' states can either be connected ($7$ and $10$) or disconnected ($7b$ and $10b$). Using either the original case or its corresponding `b' state will yield a closed curve. To determine which to use, we calculate the density at the center of the square - if it is `in' then we assume that the `in' states are linked otherwise we assume the `out' states are linked (the `b' states).   }
  \label{fig:2}
\end{figure}

There is a caveat to the method - there is an ambiguity regarding the cases $N_{\rm c} = 7$ and $N_{\rm c} = 10$. In these cases each edge of the square will have a vertex belonging to the excursion boundary and we can join these vertices in two different ways. In Figure \ref{fig:2} we exhibit the ambiguity, labeling the squares $N_{\rm c} = 7, 7b$ and $N_{\rm c} = 10, 10b$. Regardless of which of these cases we choose to adopt, the method will always yield a closed bounding perimeter. Furthermore, $N_{\rm c}=7, 10$ are rare configurations when calculating the bounding surface of  fields that are smoothed over more than a few pixel lengths. Nevertheless, one must still account for the ambiguity. We select either $7$ or $7b$  by estimating the value of $\delta$ at the center of the square, simply as the mean of the four vertices. If this value is `in' ($\delta > \sigma_{0} \nu$), then we assume that the two `in' vertices of the square belong to the same excursion region (that is, we select $N_{\rm c}=7$). Otherwise we select $7b$. We perform a similar operation for the case $10$. 

\begin{figure}
  \begin{center}
  \includegraphics[width=0.225\textwidth]{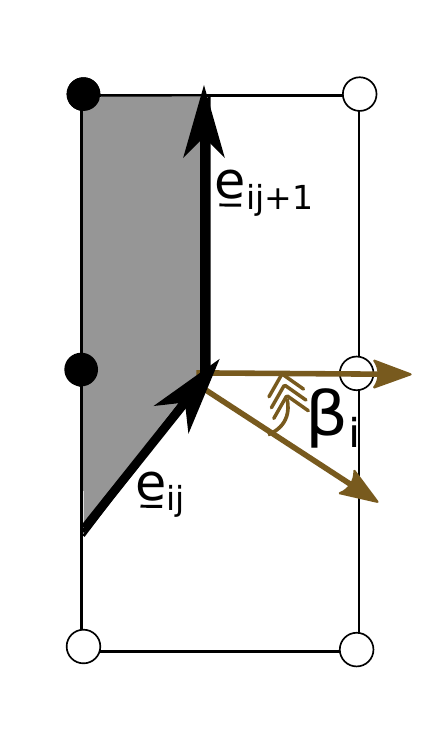}
  \end{center}
  \caption{We exhibit two squares $(i,j)$ and $(i,j+1)$, which represent cases $N_{\rm c} = 2$ and $N_{\rm c} = 6$ in Figure \ref{fig:1} - by interpolation we generate the black arrows which are the vectors ${\bf e}_{ij}$ and ${\bf e}_{ij+1}$. The light brown arrows denote the unit normals to ${\bf e}_{ij}$ and ${\bf e}_{ij+1}$ - the angle between them is $\beta_{i}$. We note that $\beta_{i}$ is directed, following the anti-clockwise conventions of our method.  }
  \label{fig:3}
\end{figure}

Once we have generated the vertices that define the bounding perimeter of the excursion set, then we can calculate its total length and enclosed area. These two quantities are proportional to Minkowski Functionals. The final Minkowski Functional in two dimensions is the genus - this can be generated by first calculating the normals to the bounding perimeter. Then, the genus is linearly related to the sum of angles between normals of adjacent perimeter edge sections. Specifically, we define the three Minkowski Functionals $W_{0,1,2}$ as

\begin{eqnarray} \label{eq:W0} & &  W_{0} = {1 \over A} \int_{Q} da = {1 \over A} \sum_{n} |\Delta A_{n}| \\ 
\label{eq:W1} & &  W_{1} = {1 \over 4A} \int_{\partial Q} d\ell = {1 \over 4A} \sum_{e} |{\bf e}| \\ 
\label{eq:W2} & &  W_{2} = {1 \over 2\pi A} \int_{\partial Q} \kappa d\ell = {1 \over 2\pi A} \sum_{i} \beta_{i}   \end{eqnarray}

\noindent where $\int_{Q} ... da$ and $\int_{\partial Q} ... d\ell$ are integrals over the area and perimeter of an excursion set respectively and $\kappa$ is the local curvature. $|\Delta A_{n}|$ is the area of the shaded region in each pixel square and $\sum_{n}$ is the sum over all pixel squares, $|{\bf e}|$ is the length of the boundary in each square (the length of the solid black arrows in Figure $\ref{fig:1}$). The sum $\sum_{e}$ indicates the sum over all edge segments in the discrete boundary. $A$ is the total area of the two dimensional plane. The angle $\beta_{i}$ between normals of adjacent perimeter segments is exhibited in Figure \ref{fig:3} - the genus is simply the sum of all such angles. The sum $\sum_{i}$ indicates the sum over all vertices in the bounding perimeter.  The genus is a topological quantity that measures the number of connected regions minus the number of holes.

The above algorithm will allow us to calculate the Minkowski Functionals and their generalisations, the Minkowski Tensors, which will be defined in section \ref{sec:MT}. These quantities describe the global properties of the excursion set. However, the total excursion set will be composed of a set of disconnected `in' and `out' sub-regions, see figure \ref{fig:1}. To calculate the properties of sub-regions we apply a simple friends of friends algorithm to the density field - for each $\delta_{ij}$ that is inside the excursion set, we assign all points $\delta_{i\pm 1, j \pm 1}$ as belonging to the same sub-region if they are also within the excursion set, and repeat the operation iteratively on these points. The only caveat is again the cases $N_{\rm c}=7$ and $N_{\rm c} = 10$ in Figure \ref{fig:1} - if two `in' vertices are linked diagonally in the box they share (for example $\delta_{ij}$ and $\delta_{i+1, j+1}$ are inside the excursion set and $\delta_{i, j+1}$ and $\delta_{i+1, j}$ are out), then we test whether the box is $7, 10$ or $7b, 10b$ by calculating the central value of the density in the box. If the square is $7, 10$ then the diagonal `in' vertices are assumed to belong to the same excursion region, otherwise they are not assigned as friends. Note that they may still ultimately be linked via our algorithm, just not through a $7b$ or $10b$ box. 

Once we have assigned each `in' grid point to a particular excursion sub-region (there are $c_{\rm id}$ distinct sub-regions), then we can calculate the area, perimeter and genus of each one, constructing a distribution of Minkowski Functionals for each density threshold $\nu$. Furthermore, we can perform the same algorithm but instead tracking the `out' states - this will yield the properties of the individual holes in the field.

The ability of this algorithm to accurately reproduce the bounding perimeter of an excursion set decreases for structures that are poorly resolved; specifically for peaks that have size roughly equal to our pixel resolution. As cosmological density fields exhibit structure on all scales, we must be careful to check that numerical artifacts do not impact our results. In appendix \ref{sec:error} we highlight two principle sources of numerical error and attempt to quantify the size of these effects. We found that smoothing the field over more than five pixel lengths is sufficient to ensure that marching squares reconstructs the excursion set boundary of the dark matter field to better than $1\%$ accuracy, for thresholds $-4 \le \nu \le 4$.

\section{Minkowski Tensors - Definition}
\label{sec:MT}

The Minkowski Functionals are scalar quantities. In \citet{McMullen:1997,Alesker1999,2002LNP...600..238B,HugSchSch07,1367-2630-15-8-083028,JMI:JMI3331} the vector and tensor generalisations were constructed - we direct the reader to these works for details of their definition. These statistics were applied to cosmology in \citet{2001A&A...379..412B, Beisbart:2001gk,Mecke:1994ax,1538-4357-482-1-L1,2017JCAP...06..023G}. 

\begin{figure}
  \begin{center}
  \includegraphics[width=0.45\textwidth]{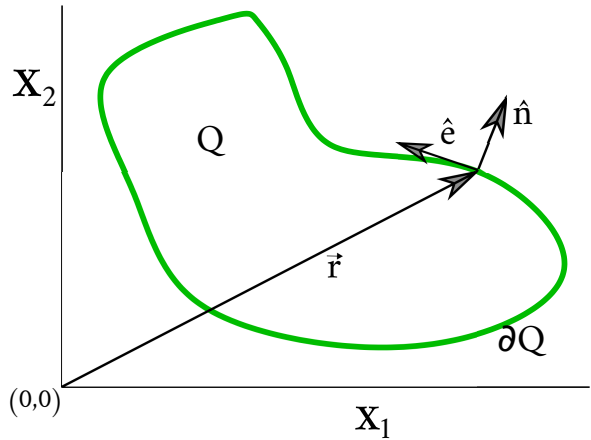}
  \end{center}
  \caption{We exhibit a schematic diagram of the area $Q$ of an excursion set, its boundary $\partial Q$ as a green solid line, and the vectors $\vec{r}, \hat{n}$ and $\hat{e}$ used in the construction of the Minkowski Tensors.}
  \label{fig:sch}
\end{figure}

The Minkowski Tensors of rank $(m,n)$ of a field in a flat two dimensional space are given by 

\begin{eqnarray} & &  W^{m,0}_{0} = {1 \over A} \int_{Q} \vec{r}^{m} da \\ 
& & W^{m,n}_{1} = {1 \over 4A} \int_{\partial Q} \vec{r}^{m} \otimes \hat{n}^{n} d\ell \\
& & W^{m,n}_{2} = {1 \over 2\pi A} \int_{\partial Q} \vec{r}^{m} \otimes \hat{n}^{n} \kappa d \ell \end{eqnarray} 

\noindent where $\vec{r}$ is the two dimensional position vector and $\hat{n}$ is the unit normal to the tangent vector of the bounding perimeter. We schematically present the vectors $\vec{r}, \hat{n}$ and $\hat{e}$ - the unit tangent vector to the boundary - in Figure \ref{fig:sch}.

The rank zero Minkowski Tensors are the standard Minkowski Functionals. Our focus is on rank two Minkowski Tensors $m+n=2$ and specifically the subset that are translationally invariant --

\begin{eqnarray}\label{eq:mt1} & & W^{1,1}_{1} = {1 \over 4A} \int_{\partial Q} \vec{r} \otimes \hat{n} d \ell \\ 
 & & W^{1,1}_{2} = {1 \over 2\pi A} \int_{\partial Q} \vec{r} \otimes \hat{n} \kappa d \ell \\ 
 & & W^{0,2}_{1} = {1 \over 4A} \int_{\partial Q} \hat{n} \otimes \hat{n} d\ell  \\
\label{eq:mt4} & & W^{0,2}_{2} = {1 \over 2\pi A} \int_{\partial Q} \hat{n} \otimes \hat{n} \kappa d \ell  
\end{eqnarray} 

\noindent Further information may be extracted from higher rank generalisations $m+n > 2$, but we do not consider these quantities in this work.

There exist relations between ($\ref{eq:mt1}-\ref{eq:mt4}$) and $W_{j} {\cal I}$, where ${\cal I}$ is the identity matrix and $W_{j}$, $(j=0,1,2)$ are the scalar Minkowski Functionals \citep{McMullen:1997} -

\begin{eqnarray} & & W_{0} {\cal I} = 2 W^{1,1}_{1} \\ 
\label{eq:We1} & & W_{1} {\cal I} = W^{0,2}_{1} + {\pi \over 2} W^{1,1}_{2} \\ 
& & W_{2} {\cal I} = 2 W^{0,2}_{2} \end{eqnarray}

These relations imply that $W^{0,2}_{2}$ and $W^{1,1}_{1}$ carry no additional information relative to the scalar Minkowski Functionals. The $W^{1,1}_{2}$ and $W^{0,2}_{1}$ tensors carry new information, with the sum being related to $W_{1}$ according to equation ($\ref{eq:We1}$). It is sufficient to measure one of these two tensors, with the other containing no new information. $W^{1,1}_{2}$ is related to $W^{0,2}_{1}$ via a rotation - $W^{0,2}_{1} = \pi T W^{1,1}_{2} T^{\rm t}/2$ where $T$ is the $\pi/2$ rotation matrix and $T^{\rm t}$ its transpose.

The tensor $W_2^{1,1}$ can be re-expressed as \citep{Chingangbam:2017uqv},

\begin{equation}
W_2^{1,1}= {1 \over 2 \pi A} \int_{\partial Q} \hat{e} \otimes \hat{e} \,d\ell,
\end{equation}

\noindent where $\hat{e}$ is the unit tangent to the curve. For a discretized field, this formula can be expressed in component form as

\begin{equation}\label{eq:w112_dis} (W^{1,1}_{2})_{ij} = {1 \over 4\pi A} \int_{\partial Q} (r_{i}n_{j}+r_{j}n_{i}) \kappa d\ell = {1 \over 2\pi A} \sum_{e} |\vec{e}|^{-1} e_{i} e_{j} \end{equation}

\noindent where $i,j$ subscripts run over the standard two dimensional $x_{1},x_{2}$ orthogonal coordinates. The sum is over all edge segments of the excursion set perimeter, $e_{i}$ is the length of the boundary segment in the $i^{\rm th}$ direction and $|\vec{e}|$ is the length of the two-dimensional vector $\vec{e}$. The diagonal components of $(W^{1,1}_{2})_{ij}$ are proportional to the (squared) total length of the excursion set bounding perimeter in the $i^{\rm th}$ direction and the off-diagonal component is the cross term. The existence of a preferred direction in the excursion boundary will manifest as an inequality between the diagonal components of $(W^{1,1}_{2})_{ij}$.

The function $(W^{1,1}_{2})_{ij}$ can be defined not only over the entire excursion set perimeter, but also over each distinct sub-region (both connected region and hole). The principle axes of each sub-region will be aligned in different directions, so to measure the shapes of these structures we extract the eigenvalues of $(W^{1,1}_{2})_{ij}$. The result is a pair of $(\lambda_{1}, \lambda_{2})$ values for each connected region and hole in the set.

We define the mean ratio of eigenvalues of all individual excursion sub-regions as 

\begin{equation} \beta_{\rm c} \equiv \left\langle {\lambda_{2} \over \lambda_{1}} \right\rangle_{\rm c} \qquad  \hspace{1.5mm}
 \beta_{\rm h} \equiv \left\langle {\lambda_{2} \over \lambda_{1}} \right\rangle_{\rm h} \qquad  \hspace{1.5mm} \beta_{\rm tot} \equiv \left\langle {\lambda_{2}\over \lambda_{1}} \right\rangle_{\rm tot}  \end{equation}

\noindent where $\langle \rangle_{\rm c,h, tot}$ denotes the sample average over all individual connected regions, holes and combined connected regions and holes respectively. $\beta_{\rm c, h, tot} \le 1$ therefore provides information regarding the mean shape of excursion regions. $\beta_{\rm c, h, tot} = 1$ corresponds to a perfectly isotropic average shape, and any value less than unity indicates some level of anisotropy - either ellipticity or a more general departure from isotropy.

Additional information is contained within $W^{1,1}_{2}$ relative to the scalar Minkowski Functionals. The statistic is invariant under translations, and a perfectly isotropic field would correspond to a diagonal matrix with equal components. Any departure from this equivalence will signify a preferred direction in the bounding perimeter of the excursion set.

\section{Applications - Two Dimensional Gaussian Random Field}
\label{sec:4}

We test our algorithm by applying it to a Gaussian random field. For such a field, the Minkowski Functionals can be calculated analytically, and we can also compare the shape of the field in the vicinity of peaks to known analytic results. 

We generate a two-dimensional Gaussian random field $\delta_{\rm k}$ in Fourier space, with constant power spectrum (Gaussian white noise). This field is then converted to its real space counterpart via a Fast Fourier Transform algorithm. We generate the field over a $3150 \times 3150 (h^{-1} \, {\rm Mpc})^{2}$ area, adopting a $2048 \times 2048$ equi-spaced grid over this range, yielding a resolution $\epsilon = 1.54 h^{-1} \, {\rm Mpc}$. We smooth the field in the plane with Gaussian kernel, using smoothing scale $R_{\rm G} = 15 h^{-1} \, {\rm Mpc}$.

\begin{figure}
  \includegraphics[width=0.45\textwidth]{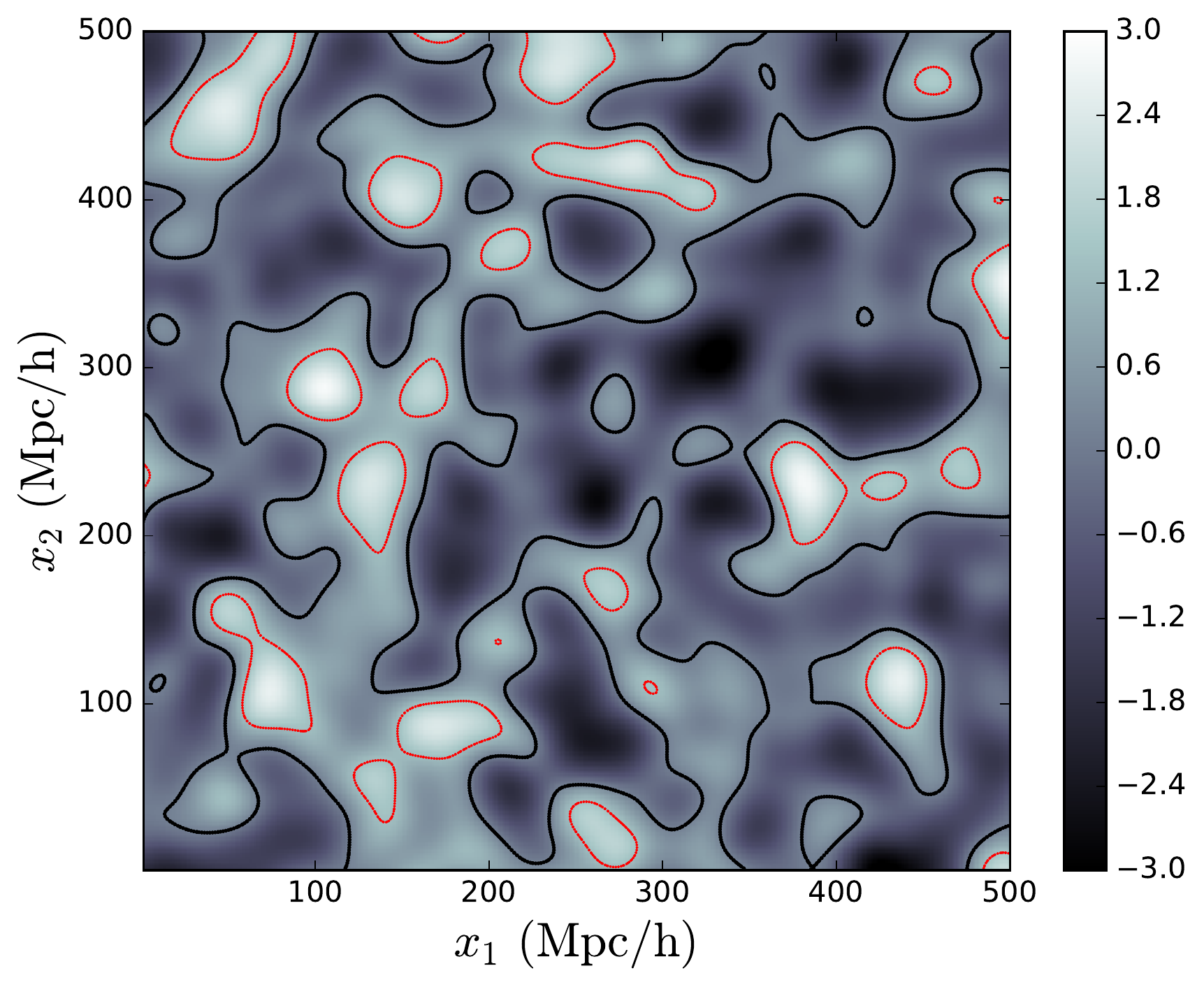}\\
  \caption{We exhibit a $500 \times 500 (h^{-1} \, {\rm Mpc})^{2}$ subset of a Gaussian random field as a heat map. The field has a flat power spectrum with a Gaussian smoothing kernel. Our algorithm generates a bounding perimeter of constant $\nu$ - we exhibit two examples $\nu=\sigma_{0}$ and $\nu = 1.4 \sigma_{0}$ as black/red contours.}
  \label{fig:gf1}
\end{figure}

We apply our two dimensional marching squares algorithm to the resulting $\delta_{ij}$. In Figure \ref{fig:gf1} we exhibit a small $500 \times 500 (h^{-1} \, {\rm Mpc})^{2}$ subset of the density field. We also exhibit an example of our algorithm - we apply a density threshold $\nu = \sigma_{0}$, $\nu = 1.4 \sigma_{0}$ and find the boundary of the excursion set. They are exhibited as black/red lines in Figure \ref{fig:gf1} -- from these boundaries we construct the Minkowski Functionals and Tensors. 

\begin{figure}
  \includegraphics[width=0.5\textwidth]{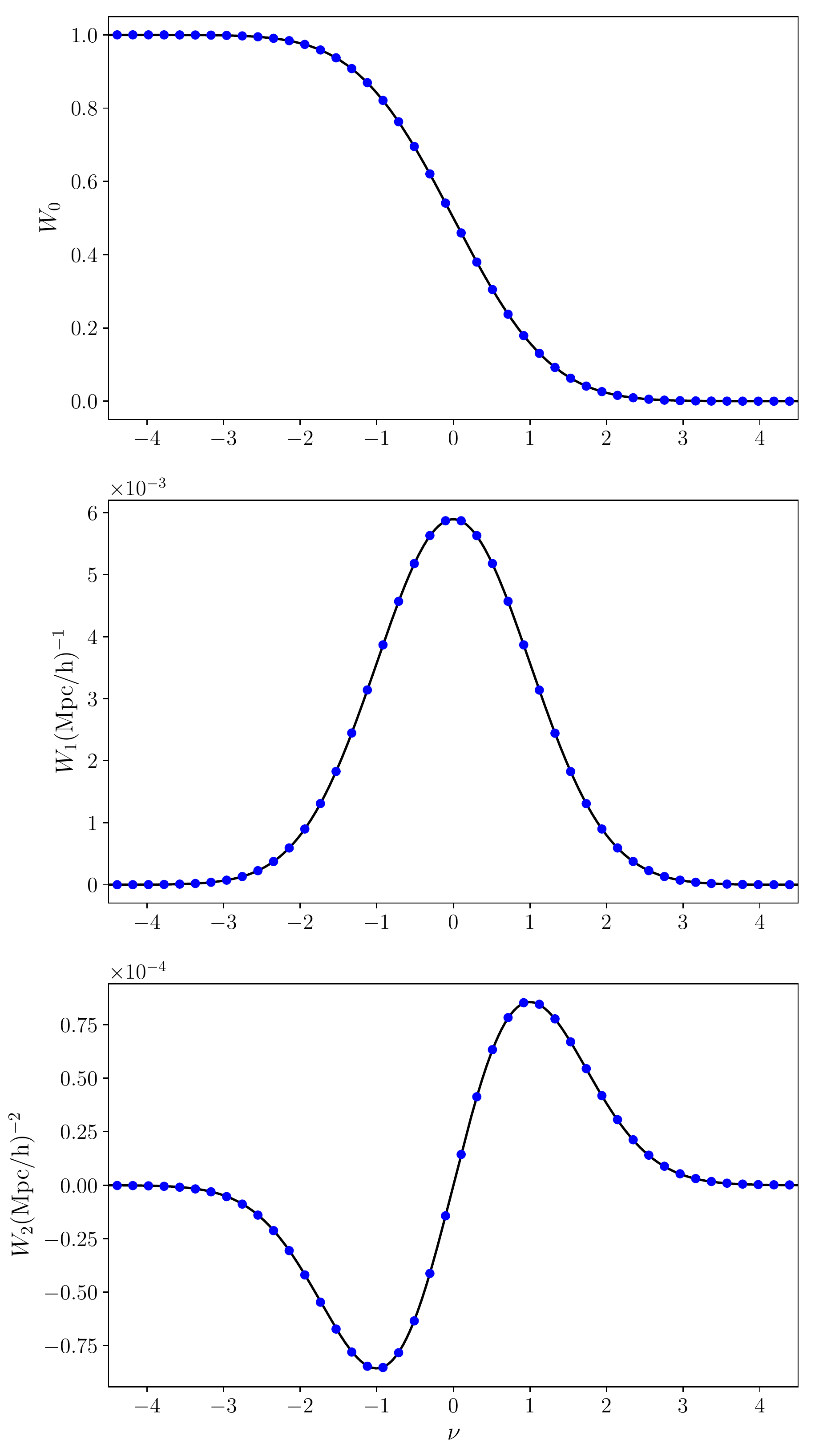}
  \caption{The Minkowski Functionals of a two dimensional Gaussian field. The blue points and error bars are the mean and error on the mean of $N_{\rm real}=100$ realisations of a Gaussian field with flat power spectrum. The solid black line is the analytic prediction. }
  \label{fig:MF2D}
\end{figure}

The scalar Minkowski Functionals are exhibited as a function of normalized density threshold $\nu$ in Figure \ref{fig:MF2D}. We generate $N_{\rm real} = 100$ realisations of a Gaussian random field - the blue points are the mean of these realisations, obtained using our algorithm. The error on the mean is smaller than the points. The solid black line is the theoretical expectation value -- the accurate reproduction of the theoretical curves serves as a consistency check of our method. 

\begin{figure}
  \includegraphics[width=0.5\textwidth]{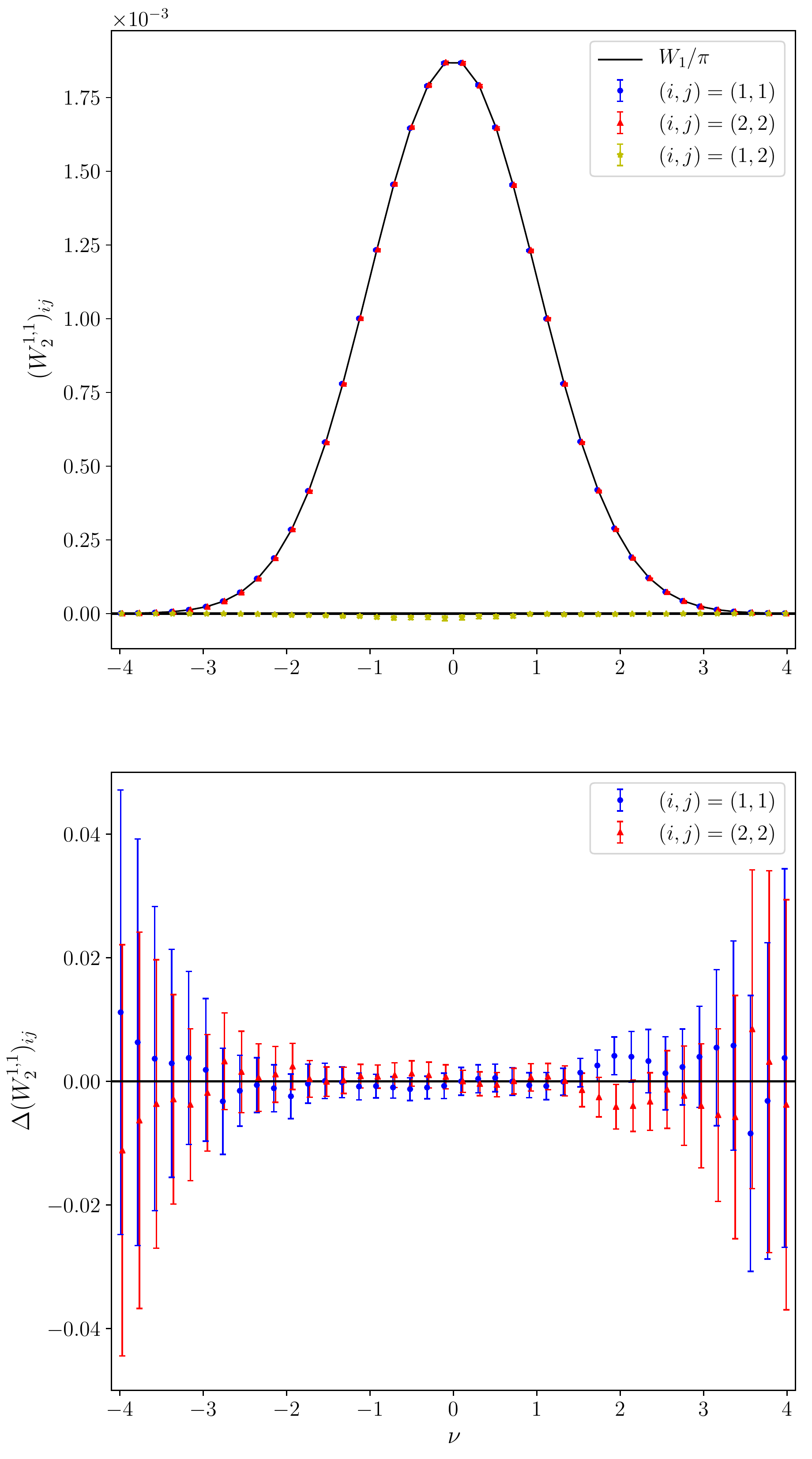}\\
  \caption{ Top panel: The matrix components of the Minkowski Tensor $(W^{1,1}_{2})_{ij}$. $(i,j)= (1,1)$ (blue), $(1,2)$ (yellow) and $(2,2)$ (red). The solid black curve is the theoretical prediction for $W_{1}/\pi$, which should match the diagonal components. Bottom panel: $(i,j) = (1,1)$ (blue) $(2,2)$ (red) components of the fractional residuals defined in equation ($\ref{eq:fres}$). These quantities are consistent with zero for all thresholds probed.}
  \label{fig:MT1}
  \end{figure}

In Figure \ref{fig:MT1} we exhibit the matrix components of $(W^{1,1}_{2})_{ij}$. In the top panel we exhibit the mean and error on the mean of diagonal components $(i,j)=(1,1), (2,2)$. We also show the theoretical prediction for $W_{1}/\pi$, which should match the diagonal components for an isotropic Gaussian random field. We find close agreement between the isotropic expectation value and our numerical reconstruction. We exhibit the off-diagonal component of the matrix, $(i,j) = (1,2)$, finding consistency with zero. 

In the bottom panel we exhibit the fractional difference 

\begin{equation}\label{eq:fres} \Delta (W^{1,1}_{2})_{ij} \equiv {\pi(W^{1,1}_{2})_{ij} - W_{1}{\cal I}_{ij} \over W_{1}} \end{equation}

\noindent which should be consistent with zero for an isotropic Gaussian field. The error bar increases with $|\nu|$ due to the smaller perimeter of the excursion set, leading to larger statistical fluctuations. Note that from the definitions in equations ($\ref{eq:W1},\ref{eq:w112_dis}$) the sum $\Delta (W^{1,1}_{2})_{11} + \Delta (W^{1,1}_{2})_{22}$ must be zero.  

In what follows we use the quantities $\Delta (W^{1,1}_{2})_{11}$, $\Delta (W^{1,1}_{2})_{22}$ and $(W^{1,1}_{2})_{12}/\langle W^{1,1}_{2} \rangle$ to study the sensitivity of the statistic $W^{1,1}_{2}$ to galaxy bias, gravitational evolution and redshift-space distortion, where $\langle W^{1,1}_{2} \rangle = [(W^{1,1}_{2})_{11}+(W^{1,1}_{2})_{22}]/2$ is the average of the diagonal components of the matrix. We note that the functions $\Delta (W^{1,1}_{2})_{11}$, $\Delta (W^{1,1}_{2})_{22}$ and $(W^{1,1}_{2})_{12}/\langle W^{1,1}_{2} \rangle$ will not be Gaussian distributed, but will be symmetric with respect to the peaks of their probability distributions.

\begin{figure}
  \includegraphics[width=0.5\textwidth]{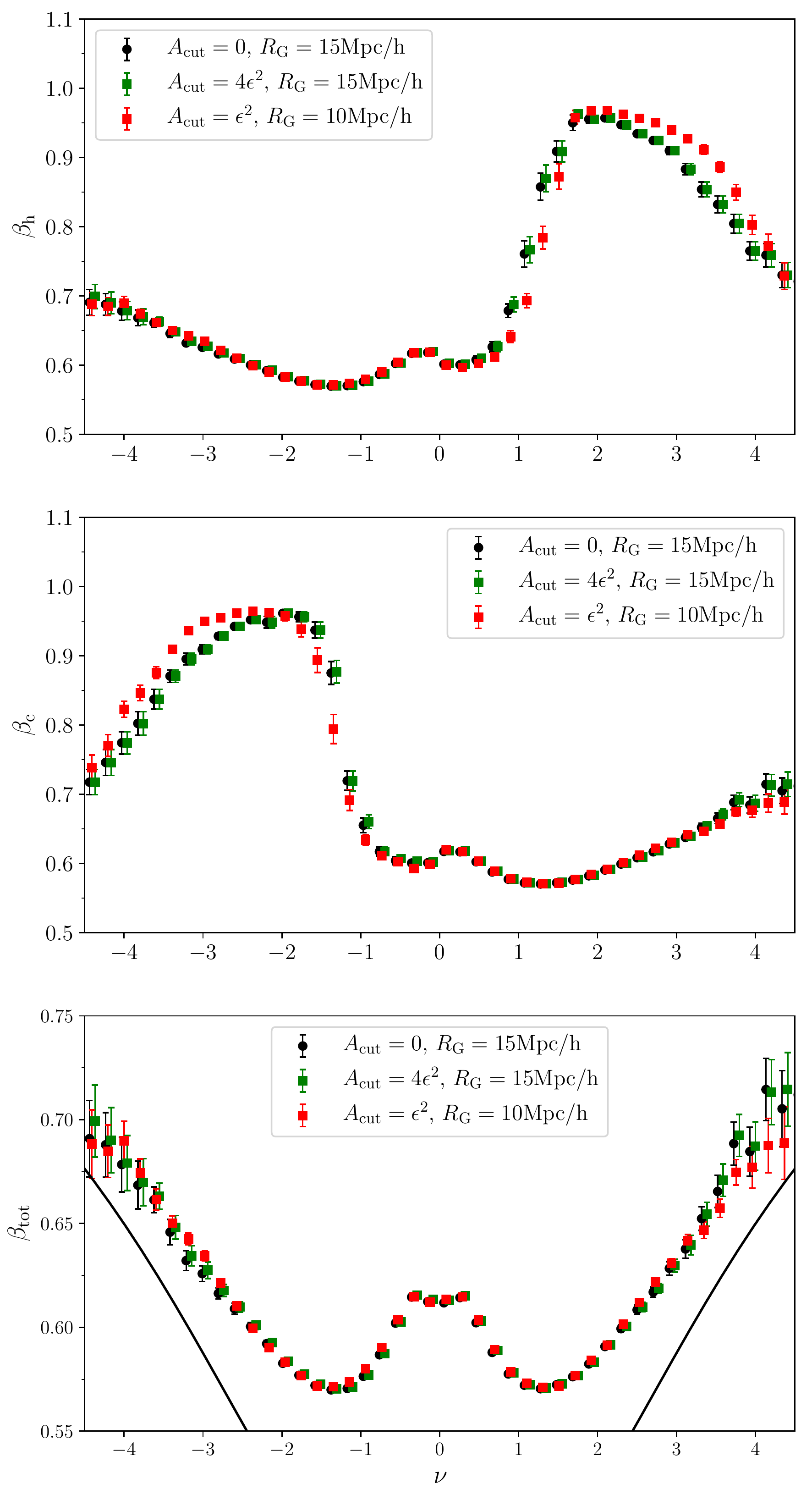}\\
  \caption{The statistic $\beta$ for holes/connected regions (top/middle panels), and the combined mean of both holes and connected excursion regions (bottom panel). The black spots and green squares indicate the mean value of this statistic after we have made different cuts to our sample of connected regions and holes (according to the size of the region), for fixed smoothing scale $R_{\rm G} =15 h^{-1} \, {\rm Mpc}$. We note that the statistic is insensitive to these cuts. The red squares are the same statistic, with a smaller smoothing scale $R_{\rm G} = 10 h^{-1} \, {\rm Mpc}$ applied. In this case the percolation of the field occurs at a slightly different value of $\nu$, but the statistics behave similarly otherwise. We also exhibit the theoretical prediction ($\ref{eq:tp}$) obtained using peak statistics as a solid black curve in the bottom panel. The theoretical curve approaches our numerical result for high threshold values.}
  \label{fig:MT2}
  \end{figure}

We exhibit the mean and error on the mean of $\beta$ in Figure \ref{fig:MT2}, generated from the $N_{\rm real} = 100$ realisations. The top/middle panels show the statistic for holes $\beta_{\rm h}$ and connected regions $\beta_{c}$ respectively, and the bottom panel shows the average for the combined holes and connected regions $\beta_{\rm tot}$. The number of distinct connected regions and holes used to calculate the averages vary greatly as a function of $\nu$ and are related to the Betti numbers \citep{2012ApJ...755..122C,Park:2013dga}. For the smoothing scales and volumes probed in this work we have $\sim {\cal O} (10^{3})$ distinct connected regions/holes at $\nu \pm 1$. The function $\beta$ is only defined in the domain $0 < \beta \le 1$, therefore it is not a Gaussian distributed variable. However, we have checked that the probability distribution function of $\beta$ is not significantly skewed and the mean is an appropriate proxy for its peak.

In Figure \ref{fig:MT2} we exhibit the statistic $\beta$ after making various cuts to the excursion set sample. To minimize spurious numerical artifacts, we have adopted a highly resolved plane of size $3150 \times 3150 (h^{-1} \, {\rm Mpc})^{2}$, with total number of pixels $N_{\rm pix} = 2048 \times 2048$ and smoothing scale $R_{\rm G} = 15 h^{-1} \, {\rm Mpc}$. With this choice we smooth over nearly ten pixels. To further test for numerical artifacts we make cuts to our sample. Specifically, the black circles and green squares in Figure \ref{fig:MT2} represent the mean $\langle \lambda_{2}/\lambda_{1} \rangle$ from a sample with $A_{\rm cut} = 0, 4\epsilon^{2}$, where $\epsilon=1.54 h^{-1} \, {\rm Mpc}$ is the pixel size and $A_{\rm cut}$ is the area cut that we apply to the excursion regions. So for the black points we use the entire sample to calculate $\langle \lambda_{2}/\lambda_{1} \rangle$, and for the green squares we remove all excursion regions (holes and connected regions) that have an area $A < 4\epsilon^{2}$ before calculating $\langle \lambda_{2}/\lambda_{1} \rangle$. As discussed in appendix \ref{sec:error}, we apply area cuts to test that no spurious anisotropic signals are generated as a result of including poorly resolved excursion subsets in the average quantities $\beta_{\rm c,h, tot}$. We find that the statistics are practically independent of any area cut that we impose, indicating the well resolved objects dominate our sample for the thresholds probed. If poorly resolved regions become dominant, then at high $|\nu|$ one would observe a spurious decrease in $\beta_{\rm c, h, tot}$. We also exhibit the same statistics for a field smoothed on a smaller scale $R_{\rm G}  = 10 h^{-1} \, {\rm Mpc}$ - we observe that $\beta_{\rm tot}$ is insensitive to $R_{\rm G}$. As discussed in appendix \ref{sec:app2} this result is expected for a Gaussian white noise field. 

The top and middle panels of Figure \ref{fig:MT2} present very different behaviour on either side of $\nu=0$ - this is due to the fact that initially we have a single hole (or connected region) with structures embedded. In this regime - the right and left hand sides of the top and middle panels respectively - the mean value of $\beta$ is dominated by a single region that is roughly the size of the entire plane\footnote{When calculating $\beta_{\rm tot}$, we do not include connected regions or holes that have size of the same order of magnitude as the total area of the plane}. This single excursion region undergoes rapid percolation into many structures, which is exhibited by the rapid change at $\nu = \pm 1$ in the figures. Following this, the statistic $\beta_{\rm h}$ is dominated by distinct holes for $\nu_{\rm A} < -1$ and $\beta_{\rm c}$ by distinct connected components for $\nu_{\rm A} > 1$.

Regardless of the $A_{\rm cut}$ that we use, the mean shape of each individual connected region and hole has value $\beta_{\rm tot} \sim 0.6$ (bottom panel) \citep{2017JCAP...06..023G}, which increases with $|\nu|$ as we expect. This suggests that the mean shape is becoming increasingly circular with increasing density threshold, but $\beta_{\rm tot}$ remains significantly smaller than unity even at large $|\nu|$. 

In the bottom panel we exhibit the theoretical prediction ($\ref{eq:tp}$) as a solid black line. The theoretical curve has been constructed by analytically calculating the mean shape of peaks of a two dimensional Gaussian field. This calculation has been performed previously in \citet{Bond:1987ub} (see also \citet{Aurich:2010qg} for later work and \citet{Bardeen:1985tr} for the three dimensional case), and we quote the results relevant to our analysis in appendix \ref{sec:app2}. In the large $\nu$ threshold limit, $\langle \lambda_{2}/\lambda_{1} \rangle_{\rm c}$ is expected to be

\begin{equation}\label{eq:tp} \langle \lambda_{2}/\lambda_{1} \rangle_{\rm c} = {\int_{0}^{2\pi} \cos^{2}\phi  \left( \kappa^{2}_{\rm m} - (\kappa^{2}_{\rm m} - 1)\cos^{2}\phi \right)^{-3/2} d\phi \over \int_{0}^{2\pi} \cos^{2}\phi  \left( 1 - (1 - \kappa^{2}_{\rm m})\cos^{2}\phi \right)^{-3/2} d\phi  }  \end{equation}

\noindent where $\kappa_{\rm m}$ is defined in equation ($\ref{eq:kap}$) and is related to the expectation value of the ellipticity of a peak $e_{\rm m}$, defined in equation ($\ref{eq:em}$). In deriving ($\ref{eq:tp}$), one assumes that contours of constant density in the vicinity of a peak are elliptical. For a Gaussian field, equation ($\ref{eq:tp}$) is valid for $\beta_{\rm tot}$ at large $\nu$, as in this regime $\beta_{\rm tot} \simeq \beta_{\rm c}$. Similarly due to the $\nu \to -\nu$ symmetry of a Gaussian field, equation ($\ref{eq:tp}$) is also valid for $\beta_{\rm h}$ at extreme negative $\nu$ values. We stress that the excursion set boundary will only trace the peaks and troughs of the field in the high $|\nu|$ threshold limit. There is no general correspondence between peaks and connected regions, or troughs and holes.

\section{Minkowski Tensors Applied to Simulated Galaxy Catalogs}
\label{sec:HR4}

We now consider the Minkowski Tensors of the low redshift dark matter density field  and study the effect of galaxy bias, gravitational evolution and redshift-space distortion on $W^{1,1}_{2}$ and $\beta_{\rm tot}$. We apply our statistics to the Horizon Run 4 simulation data. Before continuing we briefly describe the simulation. 

Horizon Run 4 is the latest data release from the Horizon Run project\footnote{http://sdss.kias.re.kr/astro/Horizon-Runs}. It is a dense, cosmological scale N-body simulation that gravitationally evolved $N=6300^{3}$ particles in a $V=(3150 \, h^{-1} \, {\rm Mpc})^{3}$ volume box. The cosmological parameters used can be found in Table \ref{tab:1}, and details of the simulation are discussed in \citet{Kim:2008kf,Kim:2015yma}. We use two dimensional slices of snapshot data at $z=0.2$, of thickness $\Delta$. The field is smoothed in the plane of the data using Gaussian kernel of width $R_{\rm G}$. We vary both $\Delta$ and $R_{\rm G}$ in what follows.

Rather than use the dark matter particle data, we adopt the mock galaxy catalog constructed in \citet{Hong:2016hsd}. Mock galaxies are assigned by the most bound halo particle-galaxy correspondence scheme. Survival time of satellite galaxies after merger is calculated by adopting the merger timescale model described in \citet{Jiang:2007xd}. We take a fiducial galaxy number density of $\bar{n} = 10^{-3} (h^{-1} \, {\rm Mpc})^{-3}$ by applying a lower mass cut. 

From the galaxy distribution we generate a density field by first generating a regular grid of size $2048 \times 2048$ in the $x_{1},x_{2}$ plane and slices of width $\Delta$ in the $x_{3}$ direction (taken as the line of sight). We bin the mock galaxies according to the $x_{3}$ slice to which they belong, and then in the two dimensional pixelated grid according to a cloud in cell scheme. Taking each slice in turn, we use the average number of galaxies $\bar{n}$ per pixel to define the two dimensional density field $\delta_{ij} = (n_{ij} - \bar{n})/\bar{n}$, where $i,j$ indices run over the $2048 \times 2048$ lattice. Next the slice is smoothed over the plane using a two-dimensional Gaussian of width $R_{\rm G}$. For each slice we calculate the Minkowski Tensor $W^{1,1}_{2}$ and $\beta_{\rm tot}$. Rather than use the conventional overdensity threshold $\nu$ to define the excursion set, instead we adopt the area threshold $\nu_{\rm A}$ parameter which is defined as   

\begin{equation}\label{eq:afrac} f_{A} = {1 \over \sqrt{2\pi}} \int^{\infty}_{\nu_{A}} \exp[-t^{2}/2] dt , \end{equation}

\noindent where $f_{A}$ is the fractional area of the field above $\nu_{A}$. The $\nu_{\rm A}$ parameterization eliminates the non-Gaussianity in the one-point function \citep{1987ApJ...319....1G,1987ApJ...321....2W,1988ApJ...328...50M}. This choice allows us to compare excursion sets in the Gaussian and non-Gaussian fields that occupy the same area.

  \begin{figure*}
  \includegraphics[width=0.5\textwidth]{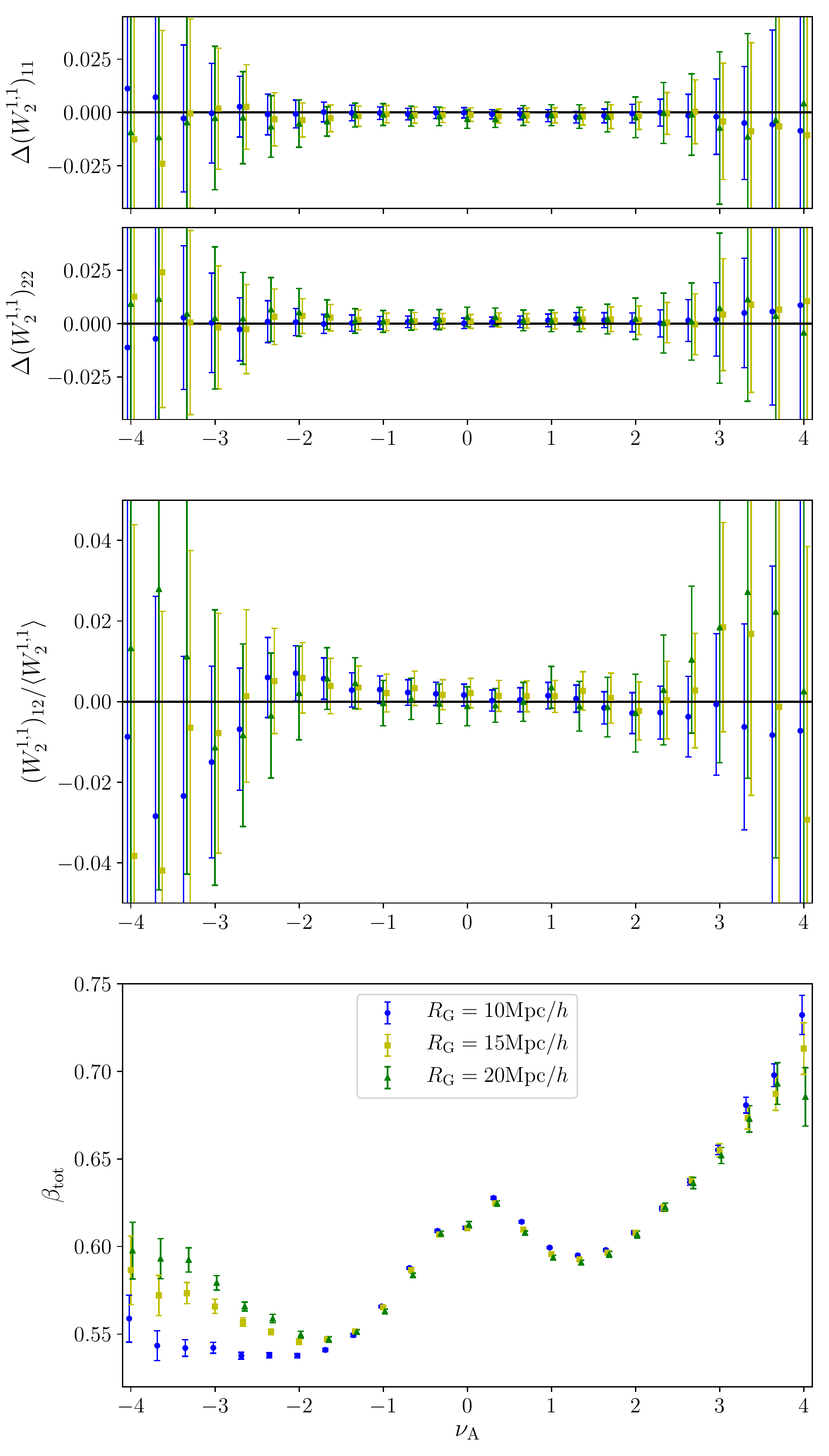}
  \includegraphics[width=0.5\textwidth]{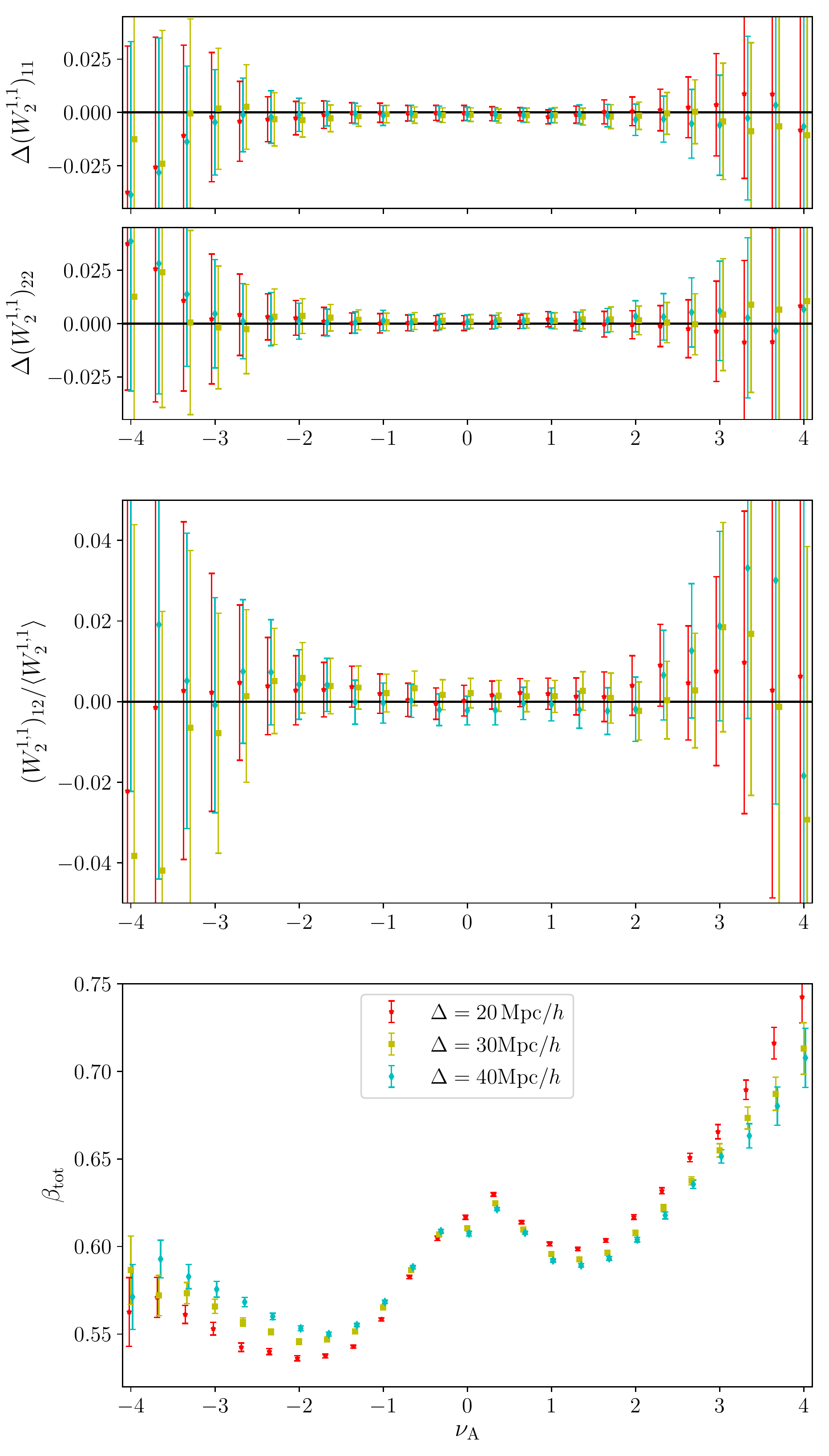}
  \caption{[Top panels] The fractional differences $\Delta (W^{1,1}_{2})_{11}$, $\Delta (W^{1,1}_{2})_{22}$ defined in equation ($\ref{eq:fres}$). We exhibit these quantities as a function of $\nu_{\rm A}$, which is related to the area fraction of the field. The dependence of $\Delta (W^{1,1}_{2})_{11}$, $\Delta (W^{1,1}_{2})_{22}$ on $\Delta$ and $R_{\rm G}$ is negligible. [Middle panels] The off-diagonal component $(W^{1,1}_{2})_{12}$ divided by the average of the two diagonal components $\langle W^{1,1}_{2} \rangle$. This quantity is consistent with zero for all $R_{\rm G}$ and $\Delta$ values. [Bottom panels] $\beta_{\rm tot}$ calculated using all connected regions and holes at each $\nu_{\rm A}$ threshold.  [Left panels] $\Delta = 30 h^{-1} \, {\rm Mpc}$, $R_{\rm G} = 20, 15, 10 h^{-1} \, {\rm Mpc}$ (green, yellow, blue). [Right panels] $R_{\rm G} = 15 h^{-1} \, {\rm Mpc}$, $\Delta = 40, 30, 20 h^{-1} \, {\rm Mpc}$ (cyan, yellow, red). Error bars denote the error on the mean from $N_{\rm slice}=75$ fields. } 
  \label{fig:HR4_lam}
  \end{figure*}

\subsection{Vary Smoothing Scales $R_{\rm G}$, $\Delta$}

In the top panels of Figure \ref{fig:HR4_lam} we exhibit the quantities $\Delta (W^{1,1}_{2})_{11}$, $\Delta (W^{1,1}_{2})_{22}$ defined in equation ($\ref{eq:fres}$) for the $z=0.2$ Horizon Run 4 mock galaxy density field. In the left panels we fix the slice thickness $\Delta = 30 h^{-1} \, {\rm Mpc}$ and vary the Gaussian smoothing scale in the plane $R_{\rm G} = 20, 15, 10 h^{-1} \, {\rm Mpc}$ (green, yellow, blue points). In the right panels we fix $R_{\rm G} = 15 h^{-1} \, {\rm Mpc}$ and vary $\Delta = 40, 30, 20 h^{-1} \, {\rm Mpc}$ (cyan, yellow, red points).  The error bars are constructed as the error on the mean calculated using $N_{\rm slice} = 75$ slices of the field. Both $W^{1,1}_{2}$ and $W_{1}$ are reconstructed from the data. In the middle panels we exhibit $(W^{1,1}_{2})_{12} / \langle W^{1,1}_{2} \rangle$. $\Delta (W^{1,1}_{2})_{11}$, $\Delta (W^{1,1}_{2})_{22}$ and $(W^{1,1}_{2})_{12} / \langle W^{1,1}_{2} \rangle$ are all consistent with zero which means that the relation $W^{1,1}_{2} \propto W_{1} {\cal I}$ remains true for the gravitationally evolved non-linear density field.

  \begin{figure*}
  \includegraphics[width=0.5\textwidth]{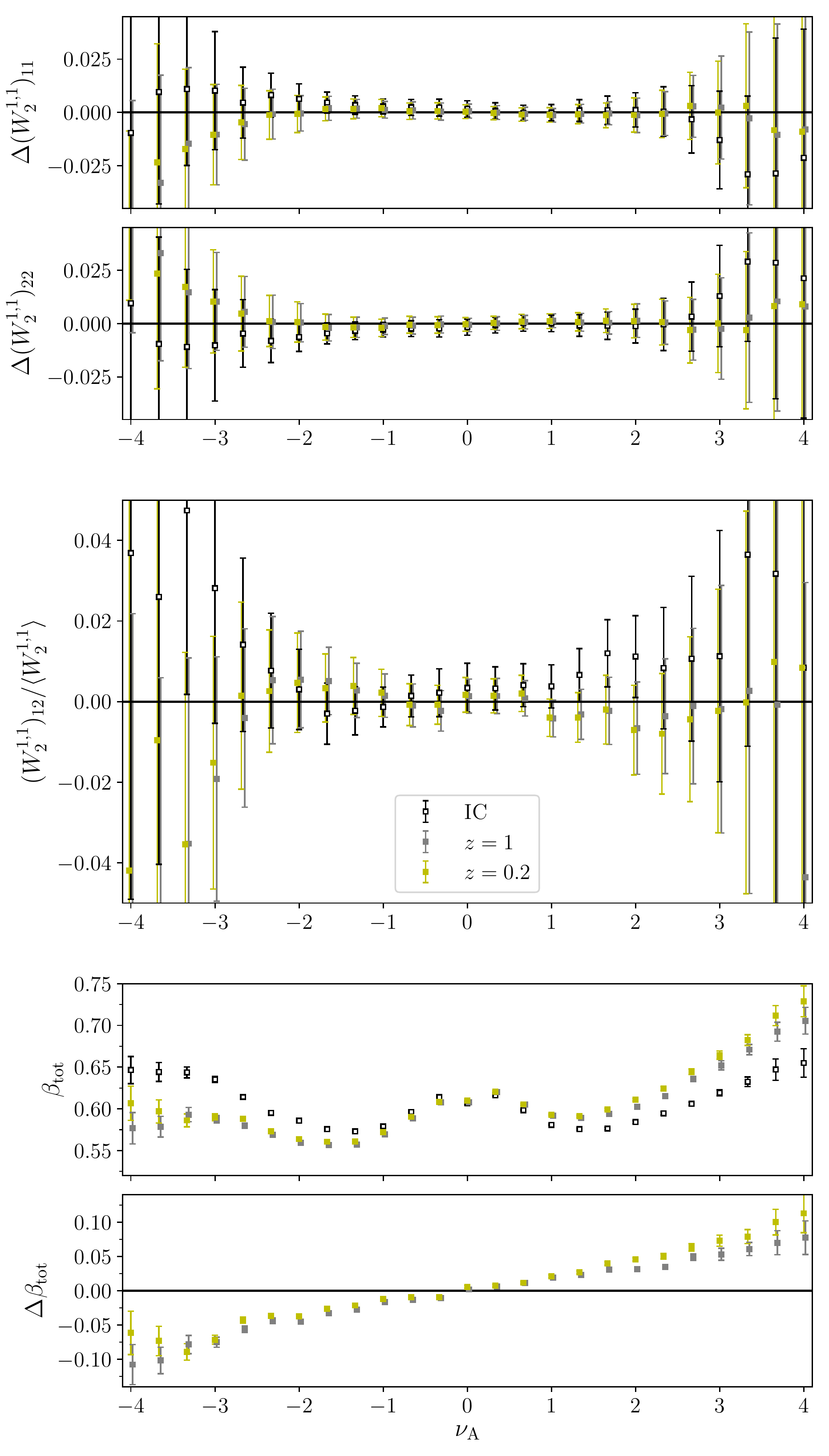}
  \includegraphics[width=0.5\textwidth]{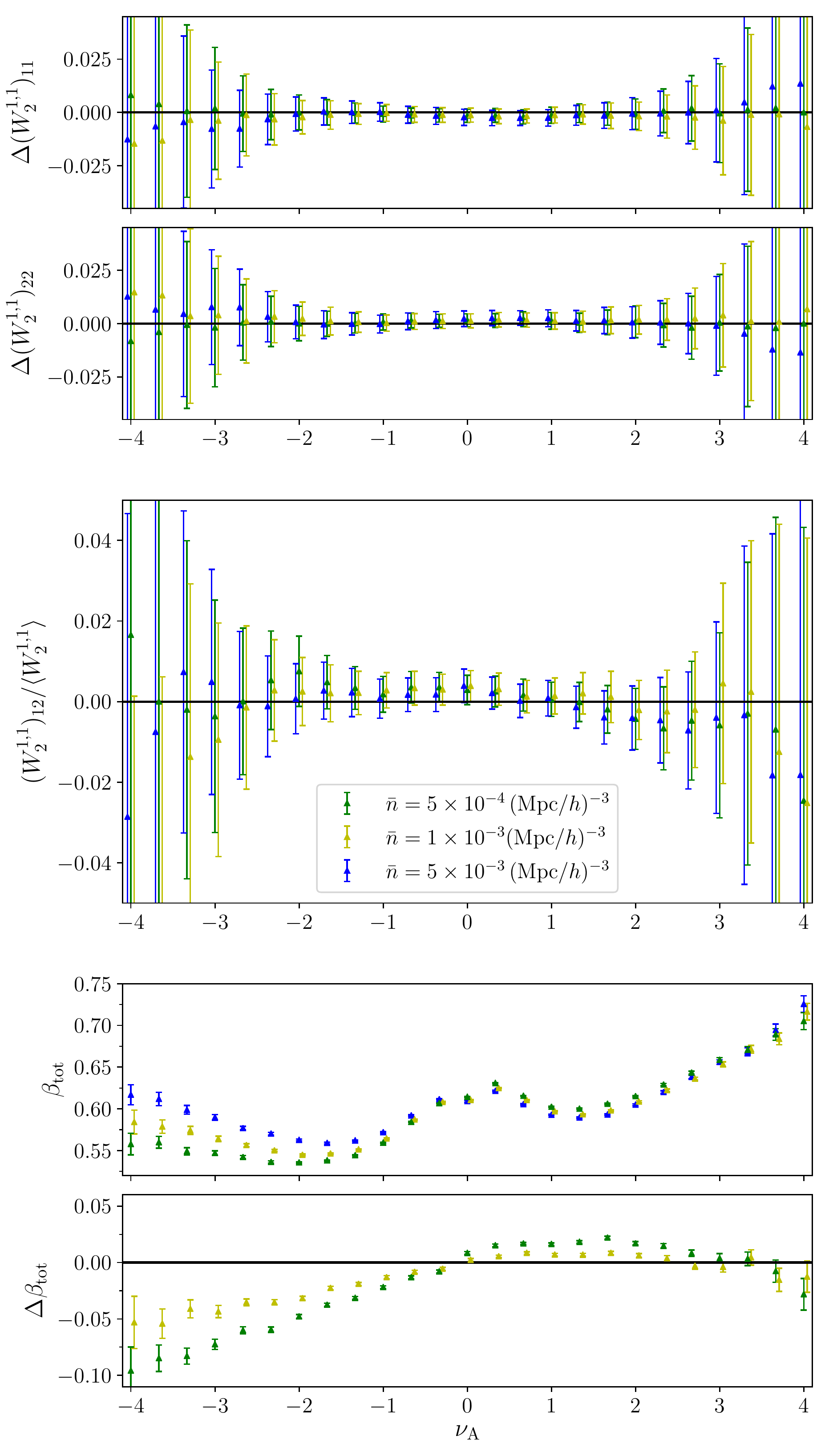}
  \caption{[Left panels] The redshift evolution of the statistics $\Delta (W^{1,1}_{2})_{11}$, $\Delta (W^{1,1}_{2})_{22}$, $(W^{1,1}_{2})_{12}/\langle W^{1,1}_{2}\rangle$ (top/middle) and $\beta_{\rm tot}$ (bottom), for the Horizon Run 4 snapshot data at $z=0.2, 1$ (yellow, grey) and the Gaussian initial condition (white). We have used fiducial smoothing parameters $\Delta = 30 h^{-1} \, {\rm Mpc}$, $R_{\rm G} = 15 h^{-1} \, {\rm Mpc}$. All galaxies in the simulation are used as density tracers at $z=1$ and $z=0.2$, with total number density $\bar{n} \simeq 1.5 \times 10^{-2} (h^{-1} \, {\rm Mpc})^{-3}$. $\Delta \beta_{\rm tot}$ is the fractional difference between $\beta_{\rm tot}$ as measured at $z=1, 0.2$ and the initial condition. [Right panels] $\Delta (W^{1,1}_{2})_{11}$, $\Delta (W^{1,1}_{2})_{22}$, $(W^{1,1}_{2})_{12}/\langle W^{1,1}_{2}\rangle$ and $\beta_{\rm tot}$ for $z=0.2$ snapshot data, taking different mass cuts to the galaxy sample to yield number density $\bar{n} = 5.0 \times 10^{-3}, 1.0 \times 10^{-3}, 5.0 \times 10^{-4} (h^{-1} \, {\rm Mpc})^{-3}$ (blue,yellow,green). $\Delta \beta_{\rm tot}$ is the fractional difference between $\beta_{\rm tot}$ measured using $\bar{n} = 1.0 \times 10^{-3}, 5.0 \times 10^{-4} (h^{-1} \, {\rm Mpc})^{-3}$ galaxy catalogs and the most dense sample $\bar{n} = 5.0 \times 10^{-3} (h^{-1} \, {\rm Mpc})^{-3}$. }
  \label{fig:evo}
  \end{figure*}

\begin{table}
\begin{center}
%\captionof{table}{}\label{tab:1} 
 \begin{tabular}{||c | c ||}
 \hline
 Parameter & Fiducial Value \\ [0.5ex] 
 \hline\hline
 $\Omega_{\rm mat}$ & $0.26$   \\ 
 \hline
 $\Omega_{\Lambda}$ & $0.74$   \\ 
 \hline
 $n_{\rm s}$ & $0.96$   \\ 
 \hline
 $\sigma_{8}$ & $0.794$   \\
 \hline
\end{tabular}
\caption{\label{tab:1}Fiducial parameters used in the Horizon Run 4 simulation.}
\end{center} 
\end{table}

We exhibit $\beta_{\rm tot}$ in the bottom panels of Figure \ref{fig:HR4_lam}. This quantity is sensitive to both $\Delta$ and $R_{\rm G}$. The most significant effect of gravitational evolution on $\beta_{\rm tot}$ is in the large $\nu_{\rm A}$ regime, where overdensities become increasingly spherical due to gravitational collapse. In contrast, underdense regions characterised by $\nu_{\rm A} < 0$ become less spherical for excursion sets of fixed $\nu_{\rm A}$. The tilt in $\beta_{\rm tot}(\nu_{\rm A})$ indicates that holes are less circular than those in a Gaussian field occupying the same area, and overdensities are more circular. $\beta_{\rm tot}$ decreases for negative thresholds $\nu_{\rm A} < 0$ as $R_{\rm G}$ is lowered, but is only weakly sensitive to $R_{\rm G}$ for $\nu_{\rm A} > 0$.

\subsection{Redshift Evolution and Galaxy Bias}

One can study the redshift evolution of $W^{1,1}_{2}$ and $\beta_{\rm tot}$ by calculating these statistics for slices of snapshot data at different redshifts. One should observe an initially symmetric $\beta_{\rm tot}$ at high redshift, which becomes increasingly tilted due to gravitational collapse with decreasing $z$. However, as we are using galaxies as tracers of the underlying field, this effect will be intertwined with galaxy bias. Fixing a constant galaxy number density at each redshift generates a galaxy distribution with a bias that is roughly constant with redshift. On the other hand, a more highly biased point distribution will better trace the high peaks of the underlying density field, which will be more spherical. Hence we can expect the tilt in $\beta_{\rm tot}$ as a function of $\nu_{\rm A}$ to increase with increasing bias. 

In the left panels of Figure \ref{fig:evo} we exhibit $\Delta (W^{1,1}_{2})_{11}$, $\Delta (W^{1,1}_{2})_{22}$, $(W^{1,1}_{2})_{12}/\langle W^{1,1}_{2}\rangle$ and $\beta_{\rm tot}$ at three epochs. We take a Gaussian random field with linear $\Lambda$CDM dark matter power spectrum as the initial condition of the simulation, and calculate $\beta_{\rm tot}$ and $W^{1,1}_{2}$ for this field and for the Horizon Run 4 snapshot boxes at $z=1$ and $z=0.2$, fixing $(\Delta, R_{\rm G}) = (30, 15) h^{-1} \, {\rm Mpc}$. We use all galaxies in the simulation, fixing the number density $\bar{n} \sim 1.5 \times 10^{-2} (h^{-1} \, {\rm Mpc})^{-3}$. 

We exhibit $\Delta (W^{1,1}_{2})_{11}$ and $\Delta (W^{1,1}_{2})_{22}$ at different redshifts, finding no evidence of evolution. Similarly the off-diagonal component $(W^{1,1}_{2})_{12}$ remains consistent with zero. This implies that the relationship $W^{1,1}_{2} \propto W_{1} {\cal I}$ is not affected by gravitational evolution. However, both $(W^{1,1}_{2})_{ij}$ and $W_{1}$ do evolve with redshift -- the scalar Minkowski functional $W_{1}$ is skewed due to the non-Gaussianity generated by the effect of gravity \citep{Matsubara:1994we,2003ApJ...584....1M}. They evolve in such a way that the relationship $W^{1,1}_{2} \propto W_{1} {\cal I}$ is preserved. 

$\beta_{\rm tot}$ becomes increasingly tilted relative to its Gaussian form with decreasing redshift. In Figure \ref{fig:evo} we exhibit both $\beta_{\rm tot}$ and the residual $\Delta \beta_{\rm tot}$, which is the fractional difference between $\beta_{\rm tot}$ as measured from the galaxy catalogs at $z=1$, $z=0.2$ and the initial condition. $\Delta \beta_{\rm tot}$ varies approximately linearly with $\nu_{\rm A}$, and is $\Delta \beta_{\rm tot} \simeq 0.1$ for high thresholds $|\nu_{\rm A}| \simeq 4$. The increasing signal with time indicates that overdense patches of fixed area become increasingly circular as collapse occurs. The underdense regions $\nu_{\rm A}<0$ of the same area become less spherical relative to the initial condition.

In the right panels we plot $\Delta (W^{1,1}_{2})_{11}$, $\Delta (W^{1,1}_{2})_{22}$, $(W^{1,1}_{2})_{12}/\langle W^{1,1}_{2} \rangle$ and $\beta_{\rm tot}$ for the $z=0.2$ snapshot data, taking different mass cuts to the galaxy distribution to generate galaxy catalogs with number density $\bar{n} = 5.0 \times 10^{-3}, 1.0 \times 10^{-3}, 5.0 \times 10^{-4} (h^{-1} \, {\rm Mpc})^{-3}$. One can observe no significant dependence of mass cut and number density on $\Delta (W^{1,1}_{2})_{11}$, $\Delta (W^{1,1}_{2})_{22}$ or $(W^{1,1}_{2})_{12}/\langle W^{1,1}_{2} \rangle$, however as we decrease the galaxy number density the value of $\beta_{\rm tot}$ decreases for $\nu_{\rm A} < 0$. We also exhibit $\Delta \beta_{\rm tot}$, which is the fractional difference between $\beta_{\rm tot}$ as measured with the $\bar{n} = 1.0 \times 10^{-3}, 5.0 \times 10^{-4} (h^{-1} \, {\rm Mpc})^{-3}$ samples and $\bar{n} = 5 \times 10^{-3} (h^{-1} \, {\rm Mpc})^{-3}$. The change in number density $\bar{n}$ affects the shape of $\beta_{\rm tot}(\nu_{\rm A})$ predominantly in the $\nu_{\rm A} < 0$ regime. One can observe that both gravitational collapse and galaxy bias affect the shape of the $\beta_{\rm tot}(\nu_{\rm A})$ curve similarly.

 \begin{figure}
  \centering
  \includegraphics[width=0.5\textwidth]{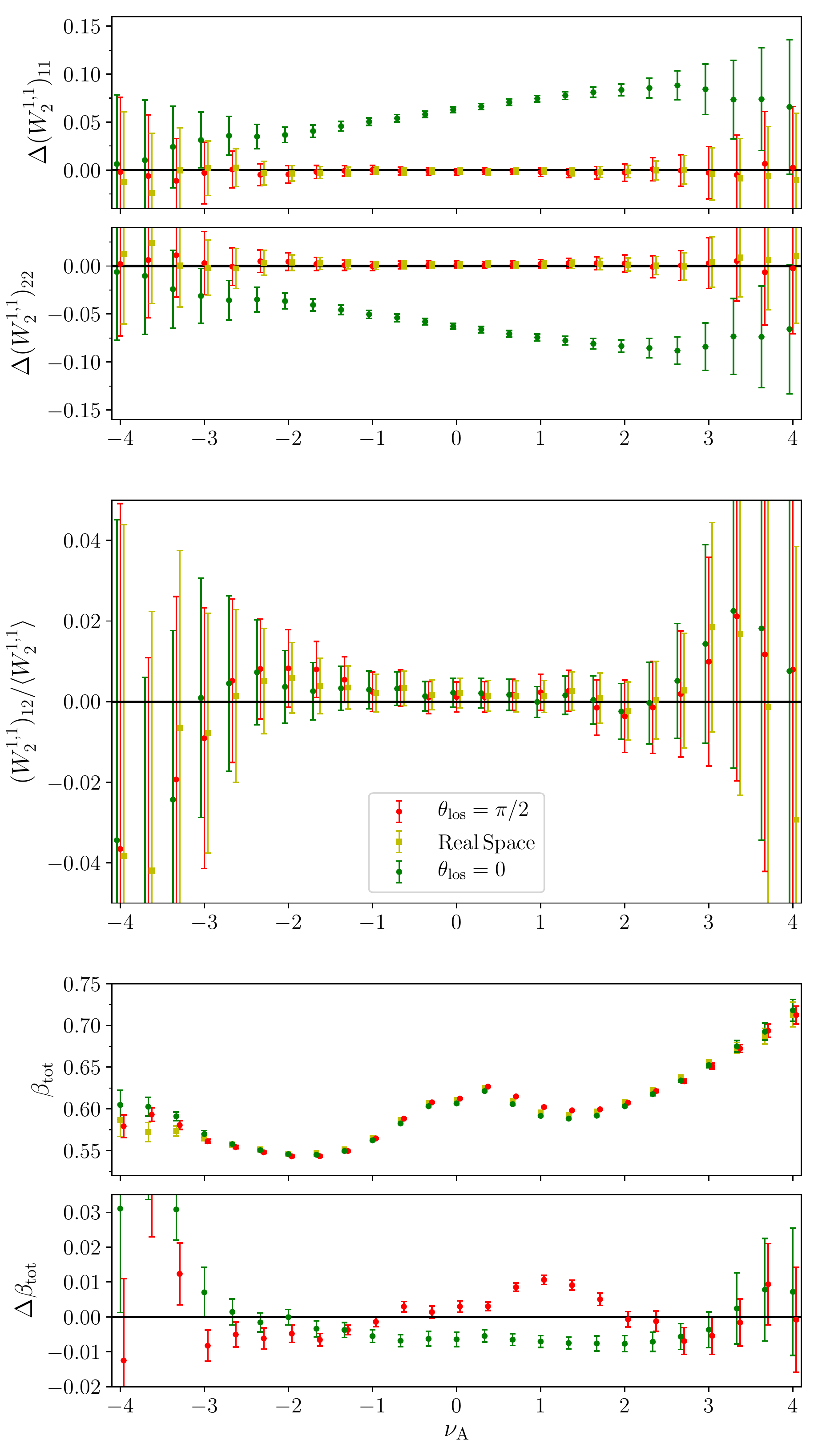}\\
  \caption{The statistics $\Delta (W^{1,1}_{2})_{11}$, $\Delta (W^{1,1}_{2})_{22}$, $(W^{1,1}_{2})_{12}/\langle W^{1,1}_{2} \rangle$ (top/middle panels) and $\beta_{\rm tot}$ (bottom panel) as a function of $\nu_{\rm A}$ for the Horizon Run 4 $z=0.2$ snapshot box, where we have introduced a redshift-space distortion along the line of sight. $\theta_{\rm los}$ is the angle of the line of sight relative to the plane of data, with $\theta_{\rm los} = \pi/2$ being the usual case where the plane is perpendicular to the line of sight. The yellow squares represent the statistics in real space, and the red/green points represent the redshift-space distorted fields perpendicular/parallel to the line of sight. The shape of individual objects $\beta_{\rm tot}$ also exhibits dependence on $\theta_{\rm los}$, with excursion subsets becoming less circular in redshift-space relative to the real space field for $\theta_{\rm los} = 0$. In the lower right panel we exhibit the fractional residual $\Delta \beta_{\rm tot}$ between $\beta_{\rm tot}$ as measured in real and redshift-space. }
  \label{fig:HR4_RSD}
  \end{figure}

\subsection{Redshift Space Distortion}

 Our results indicate that the statistic $\Delta (W^{1,1}_{2})$ is insensitive to gravitational collapse, and this remains true regardless of our choice of $\Delta$ and $R_{\rm G}$ smoothing scales. The matrix retains the relation $W^{1,1}_{2} \propto W_{1} {\cal I}$, as the effect of gravity introduces no preferred direction. However, as stated in the introduction the dark matter field that we observe via galaxy tracers is not isotropic - a preferred direction exists due to the redshift-space distortion effect along the line of sight. We close this section by considering how the global properties of the field are modified when we introduce a preferred direction to the data. For this purpose we take the $z=0.2$ snapshot data and apply a redshift-space distortion to the position of each galaxy by adjusting their position in the $x_{2}, x_{3}$ directions via the relation 

\begin{eqnarray} & &  x_{2}' = x_{2} + v_{2} {(1+z) \over H(z)} \cos \theta_{\rm los}   \\
 & &  x_{3}' = x_{3} + v_{3} {(1+z) \over H(z)} \sin \theta_{\rm los} \end{eqnarray}

\noindent where $v_{2,3}$ are the velocities in the $x_{2,3}$ directions and $\theta_{\rm los}$ is the angle of the data plane relative to the line of sight. We always generate data slices along the $x_{3}$ axis, and so varying $0 \le \theta_{\rm los} \le \pi/2$ is equivalent to varying the observer line of sight with respect to the plane. $\theta_{\rm los} = \pi/2$ is the standard case where the plane is perpendicular to line of sight, and $\theta_{\rm los}= 0$ corresponds to a density plane aligned exactly with the line of sight. The introduction of the velocity correction to galaxy positions generates a global anisotropy in the field, which the Minkowski Tensor $W^{1,1}_{2}$ is sensitive to. 

In Figure \ref{fig:HR4_RSD} we exhibit $\Delta (W^{1,1}_{2})_{11}$, $\Delta (W^{1,1}_{2})_{22}$ (top panel), $(W^{1,1}_{2})_{12}/\langle W^{1,1}_{2} \rangle$ (middle panel) and $\beta_{\rm tot}$ (bottom panel) for the real space field (yellow squares) and slices of the redshift-space distorted field aligned perpendicular (red points) and parallel (green points) to the line of sight. We fix $\Delta = 30 h^{-1} \, {\rm Mpc}$ and $R_{\rm G} = 15 h^{-1} \, {\rm Mpc}$ - these values were chosen to ensure that the field is in the mildly non-linear regime in both smoothing planes. In the bottom panel we exhibit both $\beta_{\rm tot}$ for the three cases, and also the fractional residuals $\Delta \beta_{\rm tot}$ between the real space value of $\beta_{\rm tot}$ and the redshift-space distorted values (so for example the green points in the lower panel represent the fractional residual $\Delta \beta = (\beta_{\rm tot, rsd}(\theta_{\rm los}=0) -\beta_{\rm tot, real})/\beta_{\rm tot, real}$). 

The effect of redshift-space distortion is markedly different for the two planes. If we align the data plane perpendicular to the line of sight (red points) then the effect of linear redshift-space distortion is to increase the density contrast, as galaxies in the vicinity of the slice boundary will be scattered into/out of the slice for over/under-dense regions. The shapes of connected regions and holes will change, but their orientations will remain random. As such, we can expect $W^{1,1}_{2}$ to remain insensitive to redshift-space distortion when we take $\theta_{\rm los} = \pi/2$. This agrees with our numerical result - in the top panel of Figure \ref{fig:HR4_RSD} we find no statistically significant departure of $\Delta (W^{1,1}_{2})_{11}$, $\Delta (W^{1,1}_{2})_{22}$ from zero when measured in either real or redshift-space with $\theta_{\rm los} = \pi/2$. The shape of individual excursion regions as described by $\beta_{\rm tot}$ is modified by $\sim 1\%$ but is not systematically higher or lower than its real space value.

When we align the data slice parallel to the line of sight, the effect of peculiar velocities will be to increase the ellipticity of over-densities. In contrast to the $\theta_{\rm los} = \pi/2$ case, the effect of redshift-space distortion will now generate a globally preferred direction in the excursion set boundary. The effect on individual excursion regions is small - in the bottom panel of Figure \ref{fig:HR4_RSD} we observe the fractional change $\Delta \beta_{\rm tot}$ in $\beta_{\rm tot}$ as measured in real space and the $\theta_{\rm los} = 0$ plane in redshift-space (green points). $\Delta \beta_{\rm tot}$ is negative in the range $-3 < \nu_{\rm A} < 3$ which indicates that structures in real space are more spherical, however the difference is a roughly $\sim 1\%$ effect that slowly increases with increasing $\nu_{\rm A}$.

 Although the effect on each individual excursion set region is small, it is coherent in the sense that statistically all overdensities/underdensities will be distorted in the same direction. As $W^{1,1}_{2}$ is a measurement of preferred directions in the global excursion set bounding perimeter, the distortion generates a cumulative signal in this statistic. In the top panel of Figure \ref{fig:HR4_RSD} we observe this effect - $\Delta (W^{1,1}_{2})_{22}$ exhibits a $\sim 8\%$ departure from the isotropic limit, tilting as a function of $\nu_{\rm A}$. The asymmetry of $\Delta (W^{1,1}_{2})_{22}$ about $\nu_{\rm A} = 0$ indicates that non-linear Finger of God effects, which modify the shape of overdensities parallel to the line of sight, are also contributing to the signal. The asymmetry about $\nu_{\rm A}=0$ also implies that the $W^{1,1}_{2}$ matrix no longer satisfies the relation $W^{1,1}_{2} \propto W_{1} {\cal I}$ -- non-Gaussianity of the velocity field affects $W_{1}$ and $W^{1,1}_{2}$ differently and additional information can be extracted by measuring both.

$W^{1,1}_{2}$ is sensitive to redshift-space distortion and not gravitational collapse because the latter effect is statistically isotropic, in principle at any scale. The density field will undergo collapse but no preferred direction will be generated in the excursion set boundary in real space. This makes the Minkowski Tensor an ideal candidate to measure the large scale properties of the velocity field. 

We expect that the linear Kaiser effect will generate a constant shift in $\Delta W^{1,1}_{2}$, and the Fingers of God a tilt as a function of $\nu_{\rm A}$. It follows that measurements of $W^{1,1}_{2}$ can be used to simultaneously constrain the redshift-space distortion parameter $\beta = f/b$ and the velocity dispersion of gravitationally bound galaxies. The next stage of this analysis requires a theoretical prediction of the Minkowski Tensors in redshift-space. A real space analysis has been conducted in \citet{Chingangbam:2017uqv} - the generalisation to redshift-space will be considered elsewhere.

\section{Summary} 

In this work we have studied the morphological properties of two dimensional density fields. For this purpose we have adopted the Minkowski Tensor $W^{1,1}_{2}$. To use this statistic we must first generate a bounding perimeter of constant density, which defines an excursion set. We adopted the method of marching squares, the details of which are described in the text.  We studied the diagonal and off-diagonal elements of $W^{1,1}_{2}$ for a Gaussian random field, finding that this matrix is proportional to the identity matrix and the scalar Minkowski functional $W_{1}$.

We then considered the $W^{1,1}_{2}$ statistic applied to individual subsets of the excursion set. For every threshold $\nu$ we constructed the matrix $W^{1,1}_{2}$ for each distinct connected region and hole, and from them extracted the eigenvalues $\lambda_{1,2}$. These quantities inform us of the shape of individual excursion set regions. We calculated the mean eigenvalue ratio $\langle \lambda_{2}/\lambda_{1} \rangle$ as a function of $\nu$, and in the large $|\nu|$ limit related this quantity to the mean ellipticity of the field in the vicinity of a peak. We found reasonable agreement between theory and numerical application of our algorithm in the large threshold limit. However, the statistic $\beta_{\rm tot}$ is a more general measure of shape than the ellipticity; it is a property of the excursion set boundary and makes no assumption regarding its shape.

Finally, we applied the Minkowski Tensor to mock galaxy data and considered how it is modified by gravitational collapse. We found that the mean eigenvalue ratio $\beta_{\rm tot}$ is particularly sensitive to the effect of gravity, the dominant effect being a tilt which indicates that connected regions become increasingly circular relative to holes occupying the same area. In contrast, the matrix $W^{1,1}_{2}$ defined over the entire excursion set is essentially insensitive to gravitational collapse, as the process introduces no preferred direction. 

However, when the data contains a large scale anisotropic signal, $W^{1,1}_{2}$ will exhibit strong sensitivity. When we corrected mock galaxy positions to account for redshift-space distortion and repeated our analysis using slices of the density field oriented by angle $\theta_{\rm los}$ relative to the line of sight, we found that the diagonal components of $W^{1,1}_{2}$ are significantly modified - a distinctive functional dependence on $\nu_{\rm A}$ develops with overdensities preferentially aligning along the line of sight. The anisotropy manifests as both a change in amplitude of $\Delta (W^{1,1}_{2})_{ii}$ and a roughly linear dependence on $\nu_{\rm A}$. The fact that the statistic is sensitive to anisotropy in the data, and only very weakly to the non-Gaussianity of the late time field, makes it a promising candidate to study the velocity perturbations in redshift-space. We consider the redshift-space theoretical expectation of the Minkowski Tensors in a forthcoming publication. 

To observe this signal using real data we require a measurement of the density field in planes parallel to the line of sight. Upcoming galaxy surveys such as DESI \citep{Aghamousa:2016zmz} and LSST \citep{Abell:2009aa} will provide volume limited galaxy samples over Gigaparsec volumes from which we can take subsets of the field perpendicular and parallel to the line of sight. Existing surveys such as HectoMAP \citep{2011AJ....142..133G,ASNA:ASNA201512182} provide spectroscopic galaxy catalogs over cosmological scales in slices parallel to the line of sight, and can be used to extract the redshift-space distortion signal predicted in this work. 

Photometric redshift uncertainties will be the dominant source of contamination to the signal, as they will also scatter galaxy positions along the line of sight. As the redshift-space distortion effect is present when smoothing
over large scales $R_{\rm G} \sim 15 h^{-1} \, {\rm Mpc}$, spectroscopic catalogs will be better suited to measuring $W^{1,1}_{2}$. However, upcoming photometric catalogs can still potentially be used to extract information from $W^{1,1}_{2}$. The Fingers of God introduce a tilt in $W^{1,1}_{2}$ as a function of
$\nu_{\rm A}$, as galaxies in overdense regions will predominantly experience the effect. It follows that redshift-space distortion and photometric redshift uncertainty could potentially be disentangled, as the latter will not possess the same sensitivity to density fluctuations and hence will not generate the same tilt in $\Delta W^{1,1}_{2}(\nu_{\rm A})$. Photometric redshift catalogs are characterised by large number densities and cosmological scale volumes, and using them will provide better statistics relative to spectroscopic samples. A detailed study of photometric redshift contamination will be presented elsewhere.

\acknowledgements{The authors thank the Korea Institute for Advanced Study for providing computing resources (KIAS Center for Advanced Computation Linux Cluster System) for this work.
This work was supported by the Supercomputing Center/Korea Institute of Science and Technology Information, with supercomputing resources including technical support (KSC-2013-G2-003) and the simulation data were transferred through a high-speed network provided by KREONET/GLORIAD.}

  \appendix{}

\section{Sources of Numerical Error} 
 \label{sec:error}

There are two issues with the marching squares algorithm that are capable of generating spurious numerical artifacts, which we briefly discuss.

\subsection{Topological Ambiguity Associated with Marching Squares Algorithm}

\begin{figure}[!b]
  \centering 
  \includegraphics[width=0.45\textwidth]{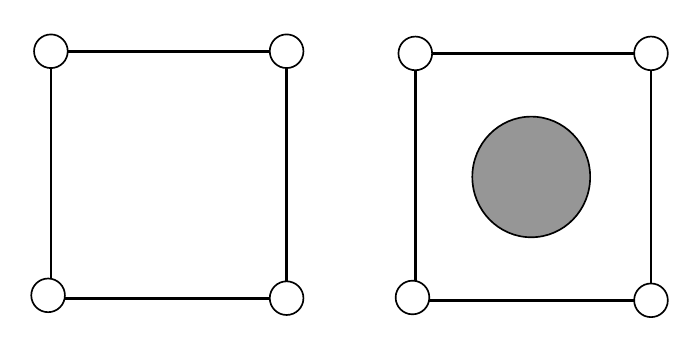}\\
  \caption{An example of the ambiguity implicit within our algorithm. Any density peak, or any non-monotonic behaviour of the field over scales smaller than our resolution $\epsilon$ will not be detected. Here we exhibit two distinct cases of a density field - in both the four vertices at which we measure $\delta_{ij}$ are `out' of the excursion region, but in the right panel there is a maxima internal to the square. Our algorithm can never detect such small scale features, and will always select the left case in this example.}
  \label{fig:num1}
\end{figure}

The first problem is our choice of interpolation scheme. We are assuming that in/out states joined along the edges of squares in the $\delta_{ij}$ grid will always cross the threshold $\delta = \nu \sigma_{0}$ once. This is tantamount to the statement that the density field is monotonically increasing or decreasing on the scale of our spatial resolution $\epsilon$. As a result, our method will not be able to distinguish certain examples of the density field - for example in Figure \ref{fig:num1} we exhibit two squares that cannot be distinguished - our algorithm will always adopt the left panel. The right panel shows a density peak internal to the square as a grey solid area - we cannot reconstruct such a peak using marching squares. Critical points are manifestly higher order phenomena, which cannot be modeled via linear interpolation.

Although our method will miss this small scale behaviour, we always smooth the field over at least three pixel lengths. The smooth field will generically be monotonic over scales $\sim {\cal O}\left(\epsilon \right)$. However, the extremes of the distribution (the high threshold peaks, for example), are likely to occupy a small area and we will inevitably fail to detect some of these objects. With the Horizon Run 4 mock galaxy data we can test the significance of this issue, using the following method. 

The mock galaxy data is a point distribution. We take the three dimensional Horizon Run 4 mock galaxies and bin them into two dimensional slices of thickness $\Delta$ as before. We then generate two grids $x_{i},y_{j}$ and $x'_{i},y'_{j} = x_{i} + \epsilon/2, y_{i} + \epsilon/2$ in the two dimensional plane and bin the galaxies according to a cloud-in-cell scheme for each grid. The resulting density fields are denoted $\delta_{ij}$ and $\delta'_{ij}$ respectively. For the density field $\delta_{ij}$ we perform the marching squares algorithm as described in the main body of the text, but now perform an additional check whenever cases $N_{\rm c} = 1$ or $N_{\rm c} = 16$ (displayed in Figure \ref{fig:1}) are encountered. 

Our algorithm will always predict $\delta < \nu$ and $\delta > \nu$ for the central values of $N_{\rm c} = 1$ and $N_{\rm c} = 16$ respectively - consistent with no small scale structures on scales $\sim {\cal O}(\epsilon)$. This is because we use a simple linear interpolation scheme to predict the density between the $(i,j)$ pixels. We now test the center of these boxes by using $\delta'_{ij}$ as the value of the density field at $x_{i} + \epsilon/2$, $y_{j} + \epsilon/2$. Whenever we encounter the case $N_{\rm c}=1$, we check if  $\delta'_{ij} > \nu$. If this inequality is satisfied, then we can say that there is some structure on the scale of our resolution $\epsilon$ that the algorithm has failed to detect. Similarly, for cases $N_{\rm c} = 16$ we test $\delta'_{ij} < \nu$ - in which case there is a hole in the excursion set that has not been detected. We count the total number of holes and connected regions that the code fails to detect in the entire plane, at each density threshold $\nu$. We then divide this number by the total number of holes and connected regions that the code successfully finds during the course of the algorithm - we denote this fraction $f_{\rm missed}$. Efficacy of the method requires $f_{\rm missed} \ll 1$. 

 \begin{figure}[b]
  \centering
  \includegraphics[width=0.45\textwidth]{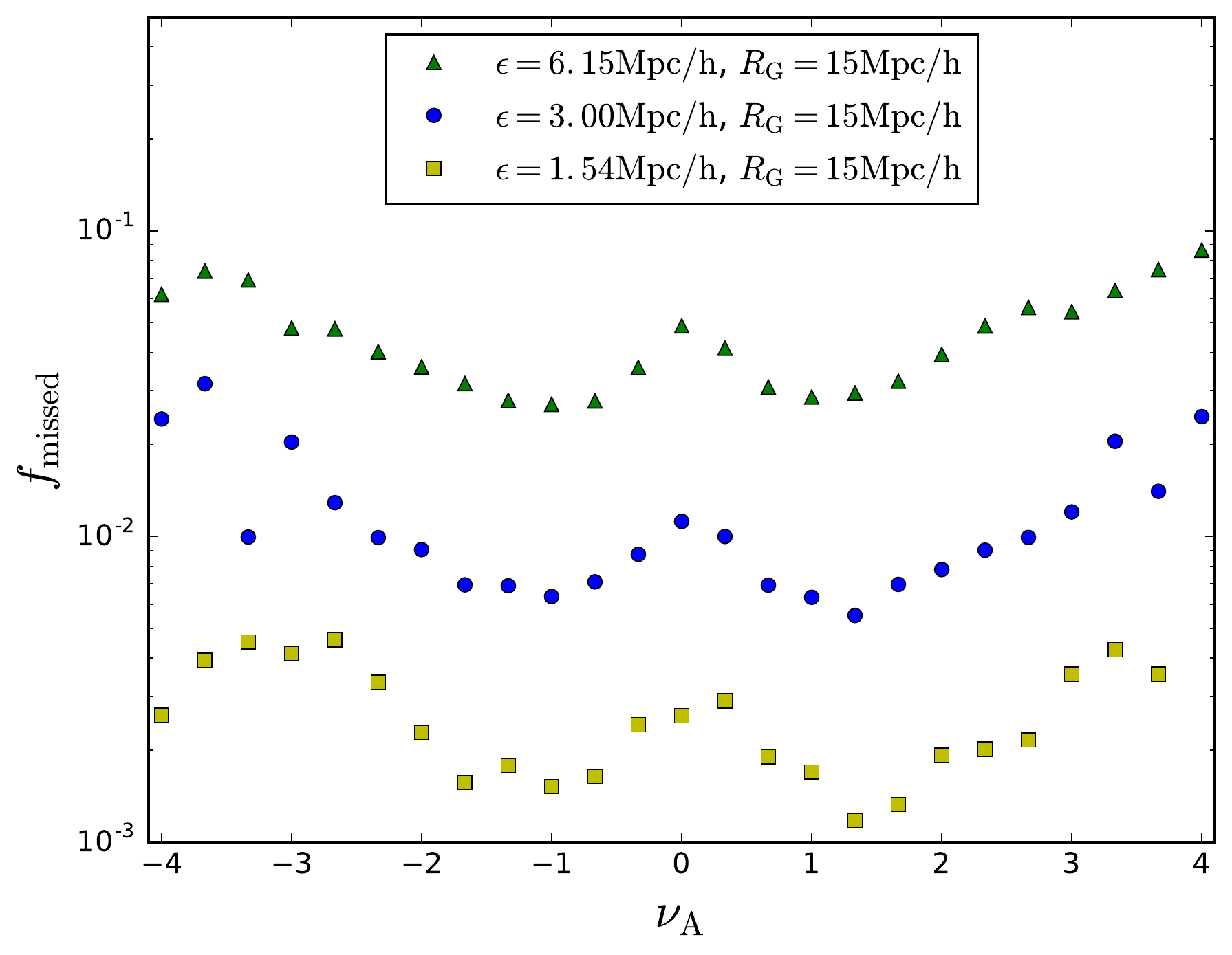}\\
  \caption{The fraction of missed connected regions and holes $f_{\rm missed}$ as a function of $\nu_{\rm A}$ for three different box resolutions $\epsilon = 1.54, 3.00, 6.15 h^{-1} \, {\rm Mpc}$ (yellow squares, blue points and green triangles). This statistic informs us of spurious numerical error in our reconstruction of the genus due to the marching squares algorithm. The number of missing structures is negligible for our choice of residual resolution $\epsilon = 1.54 h^{-1} \, {\rm Mpc}$, but increases sharply with increasing $\epsilon$.}
  \label{fig:top_amb}
  \end{figure}

We repeat this calculation for $N_{\rm slice} = 75$ slices of the three-dimensional density field, and calculate the average fraction $f_{\rm missed}$ as a function of $\nu_{\rm A}$. We exhibit this quantity in Figure \ref{fig:top_amb}. We repeat the calculation for three different spatial resolutions $\epsilon = 1.54, 3.00, 6.15 h^{-1} \, {\rm Mpc}$ (yellow squares, blue points and green triangles), fixing the smoothing scale $R_{\rm G} = 15 h^{-1} \, {\rm Mpc}$ in the plane. We observe that the fraction of missed structures is negligible for our fiducial resolution $\epsilon = 1.54 h^{-1} \, {\rm Mpc}$, but the effect increases with $\epsilon$. Additionally, the fraction of missed structures increases with increasing $|\nu_{\rm A}|$. This is to be expected, as the typical area occupied by peaks and minima decreases at large thresholds. The total number of missed holes and connected regions is between $5-10\%$ when using $\epsilon = 6.15 h^{-1} \, {\rm Mpc}$, corresponding to smoothing over $2.4$ pixels. We must smooth over at least five pixel lengths to ensure that the number of structures missed by the algorithm is $\sim 1\%$. 

It is difficult to provide a physical interpretation of $f_{\rm missed}$, as our test does not reveal all cases in which the algorithm can fail. For example we have only calculated the density field in the exact centre of the squares - peaks of the field may occur at any point. Furthermore all sixteen cases in Figure \ref{fig:1} can exhibit non-linear behaviour of the field on scales of order $\epsilon$ which can modify the genus, and we have only considered failures associated with $N_{\rm c} = 1,16$. However, we can argue that the center of the boxes $N_{\rm c} =1,16$ are most likely to exhibit irregularities (being maximally distant from our interpolation points), and hence $f_{\rm missed}$ provides a conservative indicator of the failure rates in all boxes. We conclude that the fiducial smoothing and resolution scales adopted in this work are sufficient to minimize this particular spurious numerical artifact.

\subsection{Finite Resolution Effect}

\begin{figure}
  \centering
  \includegraphics[width=0.45\textwidth]{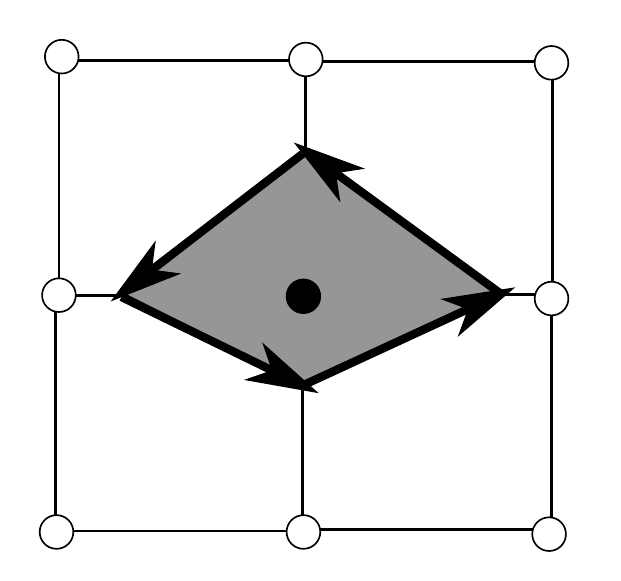}\\
  \caption{An example of the boundary that the marching squares algorithm will generate for a small excursion set region, of order of the size of a single pixel $\sim {\cal O} \left(\epsilon^{2} \right)$. The discrete nature of our algorithm generates a polygon that will not accurately represent the true, smooth boundary.}
  \label{fig:num2}
\end{figure}

A second source of numerical contamination arises for excursion set regions occupying an area of the order of the pixel size, as the reconstructed bounding perimeter for these objects will not accurately represent the smooth contour of the continuous field $\delta(x_{1},x_{2})$. As an example in Figure \ref{fig:num2} we exhibit an excursion set represented by a single point in our grid. The boundary is exhibited as a polygon, but the difference between the discretized perimeter and the smooth underlying field is likely to be large in this instance. This difference will lead to large errors in our numerical reconstruction of the morphological properties of the field whenever small excursion set regions dominate the total excursion region. This is likely to occur at large threshold values $|\nu|$. 

\begin{figure}
  \centering
  \includegraphics[width=0.45\textwidth]{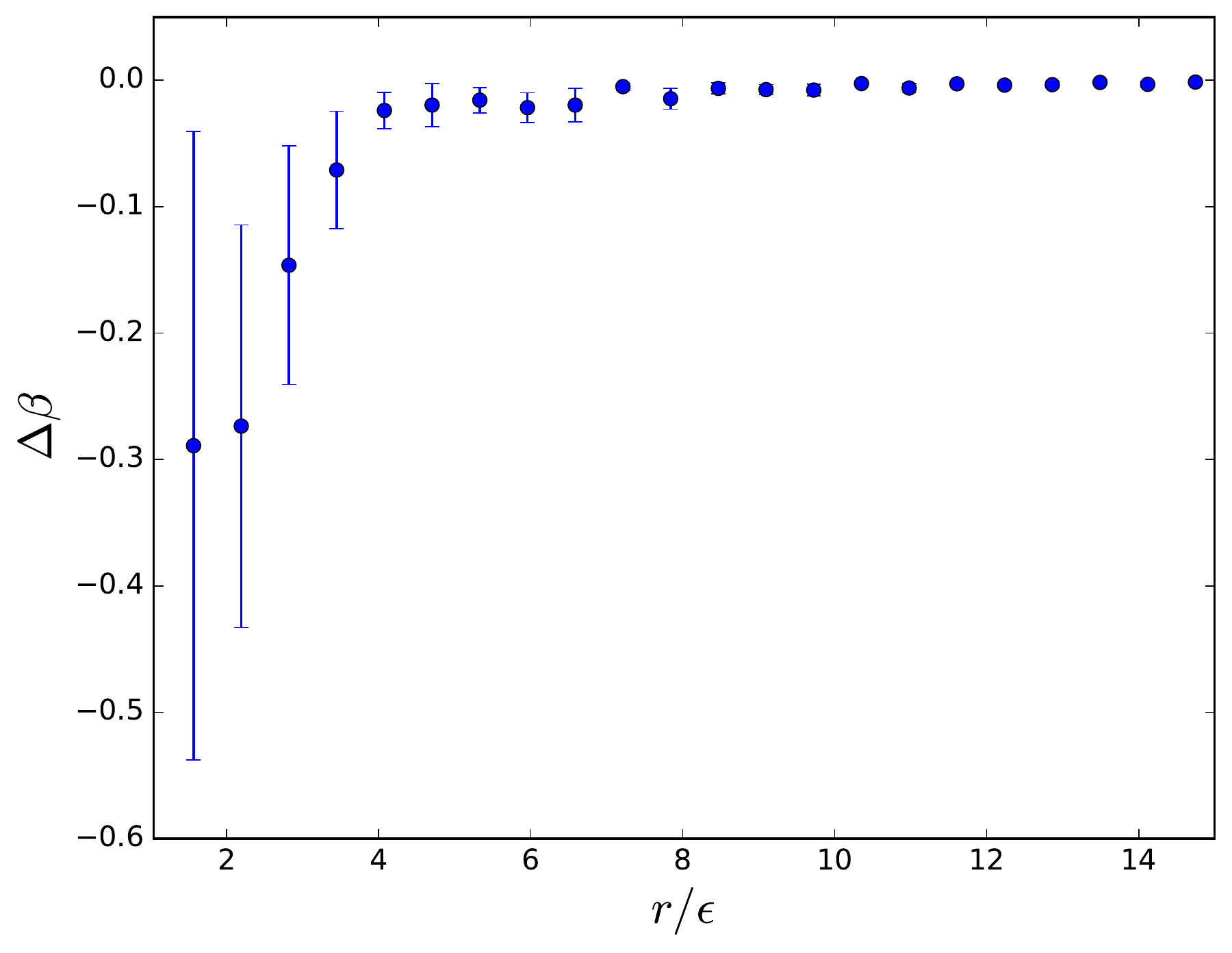}\\
  \caption{The fractional departure of the statistic $\beta$ from its theoretical expectation value $\beta=1$ for a single circular density region, as a function of the ratio of the radius of the circle to the resolution of our two dimensional grid. For circular regions that are not well resolved $r < 3\epsilon$, we find error $> 10\%$ in our numerical reconstruction of the statistics. }
  \label{fig:circles}
\end{figure}

We can estimate the magnitude of this discretization effect directly. To do so, we take a regular grid and generate a smooth circular density field. Defining the center of a circle ${\bf r}_{\rm cen} = (x_{\rm 1, cen}, x_{\rm 2, cen})$, we define $\delta(x_{1},x_{2})$ as  

\begin{equation} \delta(x_{1},x_{2}) = {\delta_{\rm cen} \over 1 + \sqrt{ (x_{1} - x_{\rm 1, cen})^{2} + (x_{2} - x_{\rm 2, cen})^{2}}} \end{equation} 

\noindent where $\delta_{\rm cen}$ is an arbitrary constant. For this field, surfaces of constant $\delta(x_{1},x_{2}) = \delta_{\rm c}$ will generate an excursion set of constant radial distance from ${\bf r}_{\rm cen}$. Increasing the threshold $\delta_{\rm c}$ will generate smaller circular excursion sets. For a circle we can trivially calculate all properties of the excursion set, and specifically we have $\beta_{c} = 1$. Therefore as we decrease the radius of the circular density field, we can ascertain the extent to which our numerical reconstruction of $\beta$ deviates from unity. In Figure \ref{fig:circles} we exhibit the residual $\Delta \beta = \beta - 1$ as a function of the radius of the circular region in units of pixel size - $r/\epsilon$. The blue points and error bars denote the average and rms fluctuations of $N_{\rm real} = 100$ realisations, randomly placing the center of the circular density ${\bf r}_{\rm cen}$ within the box. 

One can observe close agreement between theory and numerical approximation when the circle is well resolved $r > 4\epsilon$, but order $\sim {\cal O}(10\%)$ discrepancies are apparent for $r \sim \epsilon$. This numerical artifact will artificially decrease the isotropy of the density peaks. However, when we calculate $\beta_{\rm c,h,tot}$ for a stochastic density field we should not simply remove poorly resolved excursion set regions from the sample and calculate the mean properties of the remaining set, because doing so could introduce a selection bias. The shape of an excursion set is correlated with its area - large regions are more likely to be less circular. Making size cuts to our sample will bias the resulting $\beta$ statistic. So we have two competing effects - if we simply calculate the mean statistic $\langle \lambda_{2}/\lambda_{1} \rangle$ for all regions in our sample, then we will observe a spurious anisotropic signal at high $|\nu|$ threshold values where the peaks will typically be small. However, if we make an area cut and eliminate small regions from the average, then we will be preferentially selecting large excursion regions in our sample. In the main body of the text we vary the cut and examine its effect on our statistics. As we smooth the field over an increasingly large number of pixels, this numerical artifact will become less significant, and we find the effect is negligible if we smooth over several pixel lengths.

We have repeated the above test using elliptical density fields randomly located within a two dimensional plane. We again find percent level agreement between our numerical algorithm and analytic predictions, subject only to the condition that the ellipse is well resolved (with minor axis $e > 4\epsilon$). This indicates that our numerical error will not be sensitive to the shape of the excursion set regions.

\section{Mean Shape of a Peak in a Two Dimensional Gaussian Field}
  \label{sec:app2}  

A peak in a two dimensional field $\delta(x_{1},x_{2})$ can be characterized by its height, which we define as $\nu_{\rm p}$, and its ellipticity $e$. In the vicinity of a peak at $x_{i}=0$ we can expand the density field as 

\begin{equation}\label{eq:te1} \delta(x) = \delta(0) + {1 \over 2} \sum_{ij} \zeta_{ij} x_{i} x_{j}  \end{equation} 

\noindent where $i,j$ subscripts run over the two dimensional $x_{1},x_{2}$ coordinate system. $\zeta_{ij} = \nabla_{i}\nabla_{j} \delta$ is a matrix composed of second derivatives of $\delta$. We can diagonalise $\zeta_{ij}$ and re-write ($\ref{eq:te1}$) in terms of its eigenvalues $\omega_{1,2}$ -

\begin{equation} \delta(r_{i}) = \delta(0) - {1 \over 2} \sum_{i} \omega_{i} r_{i}^{2} \end{equation} 

\noindent where $r_{i}$ is a coordinate basis in this rotated frame. We fix $\omega_{1} > \omega_{2}$ in what follows. A surface of constant $\delta(r) = \delta_{\rm c}$ corresponds to an ellipse with axes 

\begin{equation} a_{i} = \left[ {2 (\delta(0) - \delta_{\rm c}) \over \omega_{i}} \right]^{1/2} \end{equation} 

\noindent The ratio of the axes of the ellipse $a_{2}/a_{1}$ is then simply given by 

\begin{equation} {a_{2} \over a_{1}}= \sqrt{ \omega_{1} \over \omega_{2}} \end{equation} 

\noindent and we define the ellipticity as 

\begin{equation} e  = {\omega_{1} - \omega_{2} \over 2(\omega_{1} + \omega_{2}) } \end{equation}

\noindent For a Gaussian random field $\delta(x_{1},x_{2})$ with power spectrum $P_{\rm 2D}(k)$, one can calculate the conditional probability of the ellipticity $e$ of a peak of height $\nu_{\rm p}$ - it is given by \citep{Bond:1987ub}

\begin{eqnarray}\label{eq:Penu} & & P(e|\nu_{\rm p}) de = { (1 - 4e^{2}) 8 e de \over \sqrt{1 + 8(1-\gamma^{2})e^{2}}} {e^{-4x_{*}^{2} e^{2}} \over G(\gamma, x_{*})} \left[ 1 - {1 \over 2} {\rm erfc}\left({ x_{*} \over \sqrt{2(1-\gamma^{2})\left(1 + 8(1-\gamma^{2})e^{2}\right)}}\right)\right] \\
\nonumber  & & G(\gamma, x_{*})= (x_{*}^{2} - \gamma^{2}) \left[ 1 - {1 \over 2} {\rm erfc}\left({x_{*} \over \sqrt{2(1-\gamma^{2})}}\right)\right] + x_{*}(1-\gamma^{2}) {e^{-x_{*}^{2}/[2(1-\gamma^{2})]} \over \sqrt{2\pi(1-\gamma^{2})}} + \\ 
& & \qquad \qquad + {e^{-x_{*}^{2}/(3-2\gamma^{2})} \over \sqrt{3-2\gamma^{2}}} \left[ 1 - {1 \over 2} {\rm erfc}\left({x_{*} \over \sqrt{2(1-\gamma^{2})(3-2\gamma^{2})}}\right)\right] \end{eqnarray}

\noindent where $\gamma = \sigma_{1}^{2}/\sigma_{2}\sigma_{0}$, $x_{*} = \gamma \nu_{\rm p}$ and $\sigma_{0,1,2}$ are cumulants of the density field

\begin{equation} \sigma_{j}^{2} \equiv \int {k dk \over 2 \pi} P_{\rm 2D}(k)k^{2j} \end{equation}

\noindent To predict the mean ellipticity of all peaks above density threshold $\nu$, we also require the number density of peaks ${\cal N}_{\rm p}$ - this is given by \citep{Bond:1987ub}

\begin{equation} \label{eq:a2}  {\cal N}_{\rm p}(\nu_{\rm p}) d\nu_{\rm p} = A  e^{-\nu_{\rm p}^{2}/2} {d\nu_{\rm p} \over \sqrt{2\pi}} G(\gamma,\gamma\nu_{\rm p}) 
\end{equation}

\noindent where $A$ is a normalizing factor.

\noindent For all peaks above a particular threshold $\nu$, we can therefore estimate the probability distribution of $e$ as 

\begin{equation}\label{eq:pe} P(e) = A_{0} \int_{\nu}^{\infty} {\cal N}_{\rm p}(\nu_{\rm p}) P(e|\nu_{\rm p}) d\nu_{\rm p} , \end{equation} 

\noindent where $A_{0}$ is a normalisation factor. For a given $\nu$ threshold we find the most likely ellipticity $e_{\rm m}$ as the expectation value 

\begin{equation}\label{eq:em} e_{\rm m} = \int_{0}^{1/2} e P(e) de .\end{equation}

\noindent It remains to relate $e_{\rm m}$ to the axis ratio of an ellipse and then calculate the Minkowski tensor for such a shape.

The most probable ellipticity $e_{\rm m}$ is related to the most probable axis ratio $\kappa_{\rm m} \equiv (a_{1}/a_{2})_{\rm m}$ as

\begin{equation} \label{eq:kap} \kappa_{\rm m} \equiv \left({a_{1} \over a_{2}}\right)_{\rm m}  = \sqrt{ 1 - 2e_{\rm m} \over 1 + 2e_{\rm m}} \end{equation} 

\noindent For an ellipse with axes $a_{1,2}$, we can calculate the Minkowski tensor $W^{1,1}_{2}$ analytically \citep{JMI:JMI3331}

\begin{equation} W^{1,1}_{2} = {\rm diag}\left( f^{1,1}_{2}(a_{1},a_{2}), f^{1,1}_{2}(a_{2},a_{1})\right) \end{equation} 

\noindent where 

\begin{equation} f^{1,1}_{2}(a_{1},a_{2}) = {1 \over 2} a_{1}^{2}a_{2}^{2} \int_{0}^{2\pi} {\cos^{2}\phi d\phi \over \left( a_{1}^{2} - (a_{1}^{2} - a_{2}^{2})\cos^{2}\phi \right)^{3/2}} \end{equation}

\noindent Note that the matrix is diagonal only in the coordinate basis aligned with the axes of the ellipse. In this case $f^{1,1}_{2}(a_{1},a_{2})$ and $f^{1,1}_{2}(a_{2},a_{1})$ correspond to the eigenvalues of this matrix. The ratio of these eigenvalues can be compared to $\beta_{\rm tot}$ in the large $|\nu|$ limit. 

For a Gaussian white noise density field, $\gamma$ and hence $W^{1,1}_{2}$ is independent of the smoothing scale $R_{\rm G}$ and power spectrum amplitude, which are the only parameters in the analysis. When applying these statistics to a cosmological dark matter field $W^{1,1}_{2}$ will depend on both $R_{\rm G}$ and the cosmological parameters $\Omega_{\rm mat}, n_{\rm s}$ via the $\gamma$ dependence of ($\ref{eq:Penu}$) and ($\ref{eq:a2}$).

\bibliographystyle{ApJ}
\bibliography{biblio}{}

\end{document}